\documentclass[reqno,12pt,a4paper]{amsart}

\usepackage{euscript}
\usepackage{multicol}
\usepackage{amsmath}
\usepackage{times}

\usepackage{amsmath,amsfonts,amsthm,amssymb,amsxtra}
\usepackage{tikz}
\usepackage{tikz-3dplot}
\usepackage{marvosym}
\usepackage{mathrsfs}
\usepackage{amssymb}
\usepackage{amsfonts}
\usepackage{latexsym}
\usepackage{amsthm}

\def\lb{\label}

\usepackage{mathtools,xparse}

\DeclarePairedDelimiter{\oldnormaux}{\bracevert}{\bracevert}

\NewDocumentCommand{\oldnorm}{som}{%
  \IfBooleanTF{#1}
    {\oldnormaux*{#3}}
    {\IfNoValueTF{#2}
       {\oldnormaux*{\vphantom{dq}#3}}
       {\oldnormaux[#2]{#3}}%
    }%
}

\numberwithin{equation}{section}
\theoremstyle{plain}
\newtheorem{theorem}{\bf Theorem}[section]
\newtheorem{lemma}[theorem]{\bf Lemma}
\newtheorem{corollary}[theorem]{\bf Corollary}
\newtheorem{proposition}[theorem]{\bf Proposition}

\title[Spectral theory of Schr\"odinger operators ] {Spectral theory   of Schr\"odinger operators\\  with  potentials that  are  measures\\
supported  on ${\Bbb N}$}

 \author{Oleg Safronov}

\begin{document}
\maketitle
 
\begin{abstract} We discuss  spectral properties  of the one-dimensional  Schr\"odinger operator with a potential  of   the form $\sum V(n)\delta(x-n)$. 
Our main result says   that  the absolutely continuous spectum of such an operator covers   an  interval  $[\alpha^2,\beta^2]$,
if  $V\in \ell^4$ and  the Fourier  series $\sum e^{2i kn}V(n)$  is  a  function of $k$ that is square integrable  over $[\alpha,\beta]$.
We  prove that  this result is sharp by constructing examples of  potentials $V\notin\ell^2$  for which the spectrum of the  Schr\"odinger operator is   singular.
\end{abstract}

\section{Introduction}\label{S1}

In this paper, we will look at the spectral theory of very special Schr\"odinger operators on the positive  half-line ${\Bbb R}_+=[0,\infty)$
with the  Dirichlet  boundary condition at  $x=0$.
Namely,  let \[V:{\Bbb N}\to {\Bbb R}\]  be a bounded real-valued  function.    We   define $H$ to be  the operator whose quadratic form 
is
\[
\int_0^\infty |u'|^2dx+\sum_{n=1}^\infty V(n)|u(n)|^2,\qquad u\in W_0^{1,2}({\Bbb R}_+).
\]
 Let $\delta$ be the standard  
delta  function on ${\Bbb R}$. Then $H$  can be formally understood as the operator
\[
H=-\frac{d^2}{dx^2}+\sum_{n=1}^\infty V(n)\delta(x-n).
\]
If  ${\rm Im}\,k>0$, then
there is a  unique  solution of the  equation
\begin{equation}\label{1.1}
-\psi''+\sum_{n=1}^\infty  V(n)\delta(x-n)\psi=k^2 \psi,\qquad \text{Im}\,k>0,
\end{equation}
that decays  exponentially as $x\to\infty$. 
We
denote $m(k^2)=\psi'(0)/\psi(0)$.
The spectral measure associated to $H$ is the measure $\mu$ on ${\Bbb R}$ having the property
\[
\int_{\Bbb R}\frac{1}{1+t^2}d\mu(t)<\infty,
\] and uniquely defined by
\begin{equation} \label{1.2}
m(E) = A+\int_{\Bbb R}\Bigl[\frac{1 }{t-E}-\frac{t }{1+t^2}\Bigr]d\mu(t), \quad  E\notin {\rm supp}\,\mu,
\end{equation}
with   $A\in {\Bbb R}$.

We  want to find conditions on  $V$  guaranteeing   that the absolutely continuous spectrum  of $H$  is essentially supported   on  the positive half-line ${\Bbb R}_+=[0,\infty)$.
That is,  $\mu'(t)>0$   for almost every $t>0$.
One of  such  conditions is  formulated  in terms of the Fourier series
\[
\hat V(k)=\sum_{n=1}^\infty e^{ikn}V(n)
\]
whose convergence is understood in the sense of  distributions.

\begin{theorem} \label{T1} Let the interval $[\alpha,\beta]\subset {\Bbb R}_+$ be free  of  integer multiples of $\pi$.  
Assume that 
\begin{equation} \label{1.3}
\sum_{n=1}^\infty |V(n)|^4 < \infty,\qquad \text{and}\quad \int_{\alpha}^{\beta}\bigl|\hat V(2k)\bigr|^2dk<\infty.
\end{equation}
Then  the following statements are true:
\vglue4pt
\noindent\hglue16pt\hangindent=34pt\hangafter=1
 {\rm(0)} The spectrum of $H$ coincides with   the support of $\mu$ which  is $[0,\infty)\cup 
\{E_j\}_{j=1}^{N}$ where $N$ is  zero{\rm ,} finite{\rm ,} or 
infinite{\rm ,} and $E_1 < E_2 <\cdots <0$.   If $N$ is 
infinite{\rm ,} then\break $\lim_{j\to\infty} E_j =0$.

\noindent\hglue16pt\hangindent=38pt\hangafter=1
 {\rm (1)} Let $\mu_{\rm ac}(E)=f(E)\, dE$ where $\mu_{\rm ac}$ 
is the Lebesgue absolutely continuous component of $\mu$. Then

\begin{equation} \label{1.4}
\int_a^b\log[ f (E)]\, dE >-\infty,
\end{equation}
for  any interval $[a,b]\subset (\alpha^2,\beta^2)$.
\end{theorem}

 {\it Remark.}
Condition (0) is just a quantitative way of writing that the essential spectrum of $H$ is
the same as that of $H_0=-\frac{d^2}{dx^2}$.

\bigskip

\begin{theorem}\label{T2}
Assume that 
\begin{equation} \label{1.5}
\sum_{n=1}^\infty |V(n)| < \infty.
\end{equation}
 Then 
\begin{equation} \label{1.6}
\sum_{j=1}^{N} |E_j |^{1/2}  \leq \frac12\sum_{n=1}^\infty |V(n)|.
\end{equation}
\end{theorem}

The  next statement   deals  with  the    sums  of   higher  powers of eigenvalues.

\begin{theorem} \label{T4*}Let $V\in \ell^{p+1/2}({\Bbb N})$ for $p>1/2$.  Let $E_j$  be  the negative  eigenvalues  of the operator $H$.  Assume that $\|V\|_\infty<2$.
Then 
\begin{equation}\label{1.7}
\sum_{j=1}^N |E_j|^p \leq C_p\sum_{n=1}^\infty|V(n)|^{p+1/2},
\end{equation}
where 
\[
C_p=\frac{\sqrt 2\int_0^4(1-\gamma/4)\gamma^{p-3/2}d\gamma}{\int_0^1(1-\gamma)^{1/2}\gamma^{p-3/2}d\gamma}.
\]
\end{theorem}

\vglue4pt 
 {\it Remark.}
Inequalities \eqref{1.6}, \eqref{1.7}  are  analogues of the celebrated bounds of Lieb and Thirring
\cite{LT1}, \cite{LT2} for Schr\"odinger operators with   potentials $V$  that are  functions on ${\Bbb R}_+$. 
Inequality \eqref{1.6}  could be also viewed as  an analogue of   the  bound  established  by
 Hundertmark,  Lieb, and Thomas in \cite{HLT}.

\bigskip

By property (1),   for any  $V$ satisfying  the condition \eqref{1.3},   the essential
support of the a.c. spectrum of $H$ contains the interval  $[a,b]$. That is, $\mu_{\rm ac }$ gives positive weight to
any subset of  $[a,b]$ of positive  Lebesgue   measure.   This follows from \eqref{1.4} because
$f$ cannot vanish on any such set. 
This observation may be   compared   with well-known  results  involving  ``usual" Schr\"odinger
operators whose  potentials are  functions  $V\in L^p({\Bbb R}_+)$. It is known that   such operators  
have a.c. spectrum if $p= 2$ (see Deift-Killip \cite{DK}),  but the a.c. spectrum can disappear once $p>2$.
While the assumption $V\in L^p({\Bbb R}_+)$ with $p>2$ does  not,   by itself,  guarantee the presence of absolutely continuous spectrum,
there are additional conditions that can be imposed to ensure its existence (see \cite{Kil} and \cite{MNV}).   
One should also   mention  that many properties and results regarding Schr\"odinger operators have analogous statements or extensions for Jacobi matrices  (see, for instance,  \cite{KS} and\cite{LNS}).
In this sense,  one   can  find the comparison between  \eqref{1.3}   and   the conditions imposed  on the perturbation  of the free Jacobi matrix  in \cite{LNS} particulary  interesting:
any $V\in \ell^4$ obeying
\[
\sum_{n=1}^\infty \bigl(V(n)+V(n+1)\bigr)^2<\infty,
\]
satisfies  assumptions of  Theorem~\ref{T1}. 
However, the  closest    to Theorem~\ref{T1}  is   the result of R. Killip \cite{Kil},   involving the Fourier transform of the ``usual'' potential  $V\in L^3({\Bbb R}_+)$ (for  Jacobi matrices,  see also \cite{Safronov}).
While their  final outcomes  are similar,   the proofs of Theorem~\ref{T1} and  Killip's theorem  \cite{Kil} differ in two critical ways. 
First of  all,  in our case,  $\hat V$   is the sum of the Fourier series which   is a periodic  function of $k$  that does not  decay at infinity,  even if $V$ has a finite support.
Secondly,   in the proof,  one needs to link the spectral properties of $H$ to  the properties of  some Jacobi matrices.  
This   connection  forces us to  consider  intervals $[\alpha,\beta]$  that are free  of integer multiples of $\pi$.
Also note that,  once $V$ satisfies   the  condition \eqref{1.3},   it  will also satisfy  \eqref{1.3}   with   $\alpha$ and $\beta$ replaced  by   $\alpha+\pi n$ and $\beta+\pi n$. Therefore,  the a.c. spectrum of $H$   will cover all intervals $[(\alpha+\pi n)^2,(\beta+\pi n)^2]$  with $n\in {\Bbb N}$.

\bigskip

The operator $H$ could  be viewed as an operator on quantum graph  whose  vertices  are  points in ${\Bbb N}\cup \{0\}$ and the edges are
the unit intervals  connecting these verices.  It is  known that  periodic operators  defined on such  graphs  might have  infinitely
many gaps in their spectra. In particular, for any $a\neq 0$, the  spectrum of the operator 
\begin{equation}\label{H+a}
-\frac{d^2}{dx^2}+a\sum_{n=1}^\infty \delta(x-n)
\end{equation}
has infinitely many gaps situated either to the left or to the right  of  the points $(\pi n)^2$.   Even though  the absolutely  continuous spectrum is no longer ${\Bbb R}_+$, 
it is still natural to ask,  what    conditions  guarantee  that the absolutely  continuous spectrum of  the operator
\[
H_a=-\frac{d^2}{dx^2}+a\sum_{n=1}^\infty \delta(x-n)+\sum_{n=1}^\infty V(n)\delta(x-n)
\]
is the  same as that of the operator   \eqref{H+a}.

Consider  for simplicity  the case $a>0$.  
Suppose  $V$ obeys the same conditions as before:
\begin{equation} \label{1.3b}
\sum_{n=1}^\infty |V(n)|^4 < \infty,\qquad \text{and}\quad \int_{\alpha}^{\beta}\bigl|\hat V(2k)\bigr|^2dk<\infty.
\end{equation}
where the   interval $[\alpha,\beta]\subset {\Bbb R}_+$ is  free  of  integer multiples of $\pi$.
What is   the essential support of  the  absolutely continuous  spectrum  of the operator $H_a$?
To answer this question, we  define  the  open set
\[
K_{\alpha,\beta}=\bigl\{  \lambda>0\mid\,\,\,  \frac{a \sin(  \sqrt{\lambda} )  }{\sqrt{\lambda}}+2\cos(\sqrt{\lambda})\in S_{\alpha,\beta}    \bigr\},
\]
where
\[
 S_{\alpha,\beta} =\bigl\{  \lambda\in {\Bbb R}\mid\,\,\,\lambda= 2\cos k,\quad  k\in (\alpha,\beta)    \bigr\}.
\]

\begin{theorem} \label{T5b} Let $a>0$ and let the interval $[\alpha,\beta]\subset {\Bbb R}_+$ be free  of  integer multiples of $\pi$.
Assume that $V$ satisfies  conditions \eqref{1.3b}.
Let $\mu_{\rm ac}(E)=f(E)\, dE$ where $\mu_{\rm ac}$ 
is the Lebesgue absolutely continuous component of the spectral measure $\mu$ of the operator $H_a$. 
Then  
\begin{equation} \label{1.4b}
\int_{\alpha_0}^{\beta_0}\log[ f (E)]\, dE >-\infty,
\end{equation}
for  any interval $[\alpha_0,\beta_0]\subset K_{\alpha,\beta}$.
\end{theorem}

This theorem  holds  for $a<0$  with the set $K_{\alpha,\beta}$  replaced   by  the union 
\[
K_{\alpha,\beta}\cup \bigl\{  \lambda<0 \mid\,\,\,  \frac{a \sinh(  \sqrt{-\lambda} )  }{\sqrt{-\lambda}}+2\cosh(\sqrt{-\lambda})\in S_{\alpha,\beta}    \bigr\}.
\]
The  necessity  of considering this union  is related to the   fact  that the spectrum of  \eqref{H+a} with $a<0$ contains negative points.

To demonstrate the sharpness of Theorem~\ref{T5b}, we provide the following sequence of results. This approach parallels the statements in the work of   \cite{KU} by Kotani and Ushiroya
and  the  theorems  from  Sections 8 and 9  of the paper \cite{KLS} by Kiselev,  Last and Simon.

Let $V_\omega: {\Bbb N}\mapsto {\Bbb R}$ be the random potential of the  form
\begin{equation}\label{randomV}
V_\omega(n)=
\varkappa\,\omega_n\,|n|^{-\alpha},\qquad\text{where}\quad \alpha>0, \,\,\varkappa>0,
\end{equation}
and $\omega_n$  are independent  random variables  uniformly  distributed on $[-1,1]$.
For $a\in {\Bbb R}$,  define the operator $H_\omega$ on $L^2({\Bbb R}_+)$ by
\[
H_\omega=-\frac{d^2}{dx^2}+a\sum_{n=1}^\infty \delta(x-n)+\sum_{n=1}^\infty V_\omega(n)\delta(x-n).
\]
After that we  define   the set $\sigma_{\rm ess}(H_\omega)$   as the union
\[\begin{split}
\sigma_{\rm ess}(H_\omega)=\bigl\{  \lambda>0\mid\,\,\,  \frac{a \sin(  \sqrt{\lambda} )  }{\sqrt{\lambda}}+2\cos(\sqrt{\lambda})\in [-2,2]    \bigr\} \, \bigcup \\
\bigl\{  \lambda<0 \mid\,\,\,  \frac{a \sinh(  \sqrt{-\lambda} )  }{\sqrt{-\lambda}}+2\cosh(\sqrt{-\lambda})\in [-2,2]   \bigr\}.
\end{split}
\]
\begin{theorem}\label{alpha<1/2}
The set $\sigma_{\rm ess}(H_\omega)$  is the  essential spectrum of $H_\omega$.
If $0<\alpha<1/2$, then the operator $H_\omega$ has  dense pure point  spectrum in $\sigma_{\rm ess}(H_\omega)$.
\end{theorem}
Note  that  for $a\neq 0$,  the set $\sigma_{\rm ess}(H_\omega)$   has infinitely many  gaps  situated  near the points $\lambda_n=(\pi n)^2$.
This makes  the next   theorem especially interesting.
\begin{theorem}\label{alpha=1/2}
Assume that $\alpha=1/2$.  Then  the following statements are true:
\vglue4pt
\noindent\hglue16pt\hangindent=34pt\hangafter=1
 {\rm(1)}  For almsot every $\omega$,   the  spectrum of $H_\omega$  is  singular.

\noindent\hglue16pt\hangindent=38pt\hangafter=1
 {\rm (2)} If $\varkappa>2$, then  the operator $H_\omega$  has   dense pure point  spectrum  in the regions
\[
\Bigl\{\lambda\in \sigma_{\rm ess}(H_\omega)\,\mid\quad\lambda>0,\quad \varkappa\frac{|\sin(\sqrt{\lambda})|}{\sqrt{\lambda}}>2\Bigr\}
\]
and
\[
\Bigl\{\lambda\in \sigma_{\rm ess}(H_\omega)\,\mid\quad\lambda<0,\quad \varkappa\frac{|\sinh(\sqrt{-\lambda})|}{\sqrt{-\lambda}}>2\Bigr\}.
\]

\noindent\hglue16pt\hangindent=38pt\hangafter=1
 {\rm (3)}  If $a>0$  then  the operator $H_\omega$ has   dense pure point spectrum in
\[
\Bigl\{\lambda\in \sigma_{\rm ess}(H_\omega)\,\mid\quad 4- \frac{\varkappa^2 \sin^2(\sqrt{\lambda})}{\lambda}<\Bigl(\frac{a \sin(  \sqrt{\lambda} )  }{\sqrt{\lambda}}+2\cos(\sqrt{\lambda})\Bigr)^2 \Bigr\}
\]
and  singular continuous  spectrum in
\[
\Bigl\{\lambda\in \sigma_{\rm ess}(H_\omega)\,\mid\quad 4- \frac{\varkappa^2 \sin^2(\sqrt{\lambda})}{\lambda}>\Bigl(\frac{a \sin(  \sqrt{\lambda} )  }{\sqrt{\lambda}}+2\cos(\sqrt{\lambda})\Bigr)^2 \Bigr\}
\]

\noindent\hglue16pt\hangindent=38pt\hangafter=1
 {\rm (4)} Statement {\rm (3)}  ramains  valid for $a<0$ with $\sigma_{\rm ess}(H_\omega)$ replaced by $\sigma_{\rm ess}(H_\omega)\cap (0,\infty)$. Moreover,
the operator $H_\omega$ has   dense pure point spectrum in
\[
\Bigl\{\lambda\in \sigma_{\rm ess}(H_\omega)\cap (-\infty,0)\,\mid\quad 4+ \frac{\varkappa^2 \sinh^2(\sqrt{-\lambda})}{\lambda}<\Bigl(\frac{a \sin(  \sqrt{-\lambda} )  }{\sqrt{-\lambda}}+2\cos(\sqrt{-\lambda})\Bigr)^2 \Bigr\}
\]
and  singular continuous  spectrum in
\[
\Bigl\{\lambda\in \sigma_{\rm ess}(H_\omega)\cap (-\infty,0)\,\mid\quad 4+ \frac{\varkappa^2 \sinh^2(\sqrt{-\lambda})}{\lambda}>\Bigl(\frac{a \sin(  \sqrt-{\lambda} )  }{\sqrt{-\lambda}}+2\cos(\sqrt{-\lambda})\Bigr)^2 \Bigr\}.
\]
\end{theorem}

It  follows that   $H_\omega$ has   dense pure point  spectrum  in the  intervals   situated near the  endpoints of the  gaps.
If a  part of  a   spectral band   is not  covered  by the pure point   spectrum,  then  this part 
is a  subset of  the  singular  continuous spectrum.  There  are infinitely many bands  that intersect   both  pure point and singular continuous spectra.

\bigskip

\begin{center}

\begin{tikzpicture}

\draw[purple, ultra thick]  (5.7,.03) -- (6,.03);
\draw[purple, ultra thick]  (5.7,-.03) -- (6,-.03);

\draw[green,thick,-] (6,.03) -- (9,.03);
\draw[green,thick,-] (2,.03) -- (4,.03);

\draw[purple, ultra thick]  (1.7,.03) -- (2,.03);
\draw[purple, ultra thick]  (1.7,-.03) -- (2,-.03);

\draw[purple, ultra thick]  (9,.03) -- (9.4,.03);
\draw[purple, ultra thick]  (9,-.03) -- (9.4,-.03);

\draw[purple, ultra thick]  (4,.03) -- (4.4,.03);
\draw[purple, ultra thick]  (4,-.03) -- (4.4,-.03);

\draw[green,thick,-] (6,-.03) -- (9,-.03);
\draw[green, thick,-] (2,-.03) -- (4,-.03);
\draw[green, thick,-] (6,.01) -- (9,.01);
\draw[green, thick,-] (2,.01) -- (4,.01);
\draw[green, thick,-] (6,-.01) -- (9,-.01);
\draw[green, thick,-] (2,-.01) -- (4,-.01);

\draw[thin,->] (-0.5,0) -- (9.7,0);

\draw[xshift=-1cm] (2.4,0) node[circle,fill,purple, inner sep=1pt](a){};
\draw[xshift=-1cm] (1.9,0) node[circle,fill,purple, inner sep=1pt](a){};

\draw[xshift=-1cm] (6.1,0) node[circle,fill, purple, inner sep=1pt](a){};

\draw[xshift=-1cm] (6.4,0) node[circle,fill,purple, inner sep=1pt](a){};

\draw[xshift=-1cm] (5.7,0) node[circle,fill,  purple,inner sep=1pt](a){};

\draw[xshift=-1cm] (5.5,0) node[circle,fill, purple,inner sep=1pt](a){};

\end{tikzpicture}
\end{center}

\medskip

\begin{center} {\it Fig 1}.  Pure point spectrum is red.  \\ Singular  continuous spectrum is green \end{center}

\begin{corollary} Let  $\alpha=1/2$.  Then for each $R>0$,  there  is a nonepty open subinterval   of  $(R,\infty)$  in which $H_\omega$ has 
dense pure point specrum.  Moreover, for each $R>0$,  there  is a nonepty open subinterval   of  $(R,\infty)$   in which the spectrum of $H_\omega$ is singular continuous.
\end{corollary}

Finally, consider the case $\alpha>1/2$.

\begin{theorem}\label{randomac}
If $\alpha>1/2$, then the essential spectrum of $H_\omega$   is absolutely continuous.
\end{theorem}

\section{Preliminaries}

Here we discuss  determinants  of operators on a Hilbert space.
Let ${\frak S}_p$ denote the Schatten classes of operators with norm $\|A\|_p=\rm Tr(|A|^p)$.
 In particular, ${\frak S}_1$ and ${\frak S}_2$ are the trace 
class and Hilbert-Schmidt operators, respectively.

For each $A\in{\frak S}_1$, one can define 
a complex-valued function ${\rm det}\, (1+A)$ by
setting
\[
{\rm det}\,(1+A)=\prod_j( 1+\lambda_j),
\]
where $\lambda_j$ are  eigenvalues  of $A$.
Then
\begin{equation} \label{2.1}
|{\rm det}\,(1+A)|\leq \exp (\|A\|_1)
\end{equation}
and $A\mapsto {\rm det}\,(1+A)$ is continuous in the sense that
\begin{equation} \label{2.2}
|{\rm det}\, (1+A) -{\rm det}\, (1+B)| \leq \|A-B\|_1 \exp (\|A\|_1 + \|B\|_1 +1). 
\end{equation}
Let  us also mention  the following properties: 
$$
A,B\in{\frak S}_1\   \Rightarrow\ {\rm det}\, (1+A) {\rm det}\, (1+B) = {\rm det}\, (1+A+B+AB)$$
$$ AB,BA\in{\frak S}_1\ \Rightarrow\ {\rm det}\, (1+AB) ={\rm det}\, (1+BA)  $$
$$(1+A)\text{ is invertible if and only if } {\rm det}\,(1+A)\neq 0  $$
$$z\mapsto A(z) \text{ is analytic} \Rightarrow {\rm det}\, (1+A(z)) \text{ is analytic}. $$

\section{The perturbation determinant and the Jost   function}

Let us   extend  $V$ to all of ${\Bbb Z}$ by setting $V(n)=0$ for  $n\leq 0$.
This  allows  one  to consider   the operator $\tilde H$ 
defined  in the space  $L^2({\Bbb R})$  by the  quadratic  form
\begin{equation}\label{3.1}
\int_{\Bbb R}|u'|^2dx+\sum_{n=1}^\infty V(n) |u(n)|^2,\qquad  u\in W^{1,2}({\Bbb R}).
\end{equation}
Spectral   properties  of the operators $H$ and $\tilde H$ can be described in terms of  the family of 
 infinite 
tridiagonal matrices  of the  form
\begin{equation} \label{3.2}
J_k= \left( \begin{array}{cccccc} \ddots & \vdots & \vdots & \vdots & \vdots  & \vdots\\
 \cdots  & b_1 & -1 & 0 & 0 & \cdots \\
 \cdots  & -1& b_2& -1& 0 & \cdots \\
 \cdots  & 0 & -1 & b_3 & -1& \cdots \\
\vdots & \vdots & \vdots & \vdots & \vdots & \ddots 
\end{array}\right)\quad
\text{with}\quad b_n=2\cos k +\frac{\sin k}{ k } V(n),\quad \forall n \in {\Bbb Z}.
\end{equation}

Since  $V$ is  bounded, 
$\sup_n |b_n|<\infty$ so that $J_k$ defines a bounded 
operator on $\ell^2 ({\Bbb Z})$.  
 Thus, there is a one-to-one correspondence between operators  $H$ and  families of matrices $J_k$ described  by \eqref{3.2}.
Moreover, $\psi$ is a solution of \eqref{1.1} if and  only if   the  vector $u\in \ell^2({\Bbb N})$
defined  by
\[
u_n=\psi(n),\qquad \forall n\in {\Bbb N},
\]
 satisfies  the  three-term recurrence  relation:
\begin{equation} \label{3.3}
 - u_{n+1} + b_{n} u_n(x) - u_{n-1}=0,\qquad \forall n\in {\Bbb N}.
\end{equation}
 The unique solution of  \eqref{3.3} which obeys
\vglue-12pt
\begin{equation} \label{1.30}
\lim_{n\to\infty} e^{-ik n} u_n =1,
\end{equation} is called   the Jost solution .  For our purposes, we rewrite \eqref{3.3} in the form
\begin{equation} \label{1.29}
- u_{n+1} - u_{n-1} + 2 \cos (k) u_n+\frac{\sin k}{k} V(n) u_n  = 0,\qquad n\in {\Bbb N}.
\end{equation}

Observe that  the matrix $J_k$  is
``close" to the free matrix, $J$  defined  by 
\begin{equation} \label{3.6}
J = \left( \begin{array}{cccccc} 
\ddots & \vdots & \vdots & \vdots & \vdots  & \vdots\\
\dots& 2\cos k & -1 & 0 & 0 & \dots \\
\dots&  -1 & 2\cos k &- 1 & 0 & \dots \\
\dots& 0 & -1 & 2\cos k & -1 & \dots \\
\dots&  0 & 0 & -1 & 2\cos k& \dots \\
\vdots & \vdots & \vdots & \vdots & \vdots & \ddots 
\end{array}\right).
\end{equation}

Namely, $J_k-J$  is the  diagonal  matrix
\begin{equation} \label{1.6V}
J_k-J = \frac{\sin k}{k}  \left( \begin{array}{cccccc} 
\ddots & \vdots & \vdots & \vdots & \vdots  & \vdots\\
\dots& V(1) & 0 & 0 & 0 & \dots \\
\dots&  0&   V(2) &0& 0 & \dots \\
\dots& 0 & 0 & V(3) & 0& \dots \\
\dots&  0 & 0 & 0 &  V(4)& \dots \\
\vdots & \vdots & \vdots & \vdots & \vdots & \ddots 
\end{array}\right).
\end{equation}

\bigskip

The major tool in proving the theorem  is the perturbation determinant 
defined as
\vglue-12pt
\begin{equation} \label{1.28}
L(k) = {\rm det}\, \big(J_kJ^{-1}\big).
\end{equation}
Note  that
$L(k)$ is defined  
for ${\rm Im}\,k>0$ by the trace class theory of determinants as  long as 
$J_k-J$ is trace class. 
In particular, it  is well-defined when $J_k-J$ is a finite rank  operator.

\begin{theorem} \lb{T2.4} Suppose $J_k-J\in{\frak S}_1$.
\begin{itemize} 
\item[{\rm{(i)}}] $L(k)$ is analytic in ${\Bbb C}_+\equiv\{k\mid\,{\rm Im} k>0\}$.
\item[{\rm{(ii)}}] $L(k)$ has a zero at a point $k_j\in{\Bbb C}_+$ if and  only  if 
$k_j^2=E_j<0$ is an eigenvalue of~$\tilde H$.
All zeros of $L(k)$ are simple.
\item[{\rm{(iii)}}] If $J_k-J$ is a finite rank operator,   then $L(k)$   can be extended  analytically  into the set  ${\Bbb C}\setminus \{0\}$.
\end{itemize}
\end{theorem}

{\it Proof.} (i) follows from the fact  that the map $k\mapsto (J_k-J)J^{-1}$ is analytic.

\medskip

(ii) Let $k\in{\Bbb C}_+$.   Note that $k^2$  is an  eigenvalue of  $\tilde H$  if and  only if $0$ is an eigenvalue of $J_k$.
If $E=k^2$ is not an eigenvalue of $J_k$, then $J_k/J$ 
has an  inverse (namely, $J/J_k$), and therefore $L(k)\neq 0$. 
If $E_0$ is an eigenvalue, $J_k/J$ is not invertible, so 
$L(k)=0$. Finally,  eigenvalues of $\tilde H$ are simple 
by a Wronskian argument.   This  implies that, if $0$ is an   eigenvalue of $J_k$, then it is  simple.  That $L$ has a simple zero under these circumstances comes 
from the following.

Let $L(k_0)=0$. Choose  $\varepsilon>0$ for  which  $J_{k_0}$  has  only one  eigenvalue in the disk $\{z\mid \, |z|<2\varepsilon\}$.
Define 
$$P=P_k=\frac1{2\pi i}\int_{|z|=\varepsilon }(J_k-z)^{-1}dz.
$$ 
If $|k-k_0|$  is  sufficiently small,
then $P=P_{k}$ is the projection onto the eigenvector  of $J_{k}$  and  $(I-P_k)J_k^{-1}$ 
has a removable  singularity at $k=k_0$.  Define
\begin{equation} \label{2.29}
C(k) =  (I-P)J_k^{-1}+P.
\end{equation}
Then
\begin{equation} \label{2.30}
C(k) J_k= I- P+PJ_k.
\end{equation}
Also,  define 
\begin{equation}
\begin{split}
T(k) = JC(k)= \lb{2.31} 
 I- P+PJ_k+ (J-J_k)C(k)\ \\
=1 + \hbox{trace class}. \end{split}
\end{equation}
Then $T(k)$ is analytic at $k=k_0$.
Moreover, 
\begin{equation}\begin{split}
T(k) [ J_k/J] =J [ I- P+PJ_k]J^{-1}= \\
I+J [-P+PJ_k]J^{-1}.
\end{split}
\end{equation}
Thus, 
\begin{equation}\begin{split}
L(k)\cdot {\rm det}\,(T(k)) =  {\rm det}\, (1+J[-P+(PJ_k]J^{-1})= \\
{\rm det}\, (1-P +PJ_kP) =\\
 {\rm Tr}\, PJ_kP.
\end{split}
\end{equation}
Since $ {\rm Tr}\, PJ_kP$ has a simple zero at $k_0$,   the  function   $L(k)$ has a simple zero.

\medskip

(iii) The matrix elements 
of the operator $J^{-1}$  are
\[
\frac{1}{2i \sin k}e^{ik|n-m|}.
\]
Consequently, $ \frac{\sin k}{k} V J^{-1}$  is the matrix  whose  elements are   
\[
\frac{V(n)}{2i  k}e^{ik|n-m|}.
\]
Let  $\chi: {\Bbb N}\to {\Bbb R}$    be the  characteristic  funtion of the support of $V$.   
Obviously, $\frac{\sin k}{k} V \chi J^{-1}\chi $ can  be extended  analytically  into ${\Bbb C}\setminus \{0\}$.
It remains to observe that $$L(k)={\rm det}\, (I+ \frac{\sin k}{k} V \chi J^{-1}\chi ).$$
$\Box$

\bigskip

Part (iii) of Theorem~\ref{T2.4} shows  that $L(k)$, defined initially only on~${\Bbb C}_+$,
can be continued  to an essential part of   the boundary  $\partial \, {\Bbb C}_+$.

\bigskip

As we  saw in the proof of (iii),  the operator $J$  is  invertible.   At the same time,
invertibility of  $J_k$  is  less  obvious.
The next lemma not  only confirms  that $J_k$ has a  bounded inverse,  it  also  tells us  how big  the   norm   of the inverse is.

\begin{lemma}\label{LnormJ}
There is a positive  continuous  function $C(k)$ on ${\Bbb C}\setminus \pi{\Bbb Z}$
with the property
\begin{equation}\label{normJk}
\|J_k^{-1}\|\leq \frac{C(k)}{| {\rm Im}\, k^2|},\qquad \forall\,  |{\rm Im}\, k^2|>0.
\end{equation}
The function $C(k)$ is independent of $V$.
\end{lemma}

{\it Proof.}
Let
\[
J_k u=f,\qquad  \text{for}\quad u,f\in \ell^2({\Bbb Z}).
\]
Since the values on  boundary of  $[n,n+1]$ determine the   solution of the equation   $-\psi''=k^2\psi$ inside the interval,
we can find a  continuous function $\psi\in L^2({\Bbb R})$    such that 
\[
-\psi''=k^2\psi\qquad \text{  a.e.  on }\quad {\Bbb R},\quad \text{and}\quad \psi(n)=u_n,\qquad \forall n\in {\Bbb Z}.
\]
In this case,    the  function $\psi$
has the following property:
\[
\psi'(n+0)-\psi'(n-0)=V(n)\psi(n)+\frac{k}{\sin k} f(n),\qquad \forall n\in {\Bbb Z}.
\]
Using this property and  integrating by parts, we obtain    that
\[
\int_{\Bbb R}|\psi'|^2dx+\sum_{n=-\infty}^\infty V(n)|\psi(n)|^2=k^2 \int_{\Bbb R}|\psi|^2dx+\frac{k}{\sin k}\sum_{n=-\infty}^\infty f(n)\overline{\psi(n)}.
\]
Consequently,
\[
\bigl|{\rm Im}\, k^2\big| \int_{\Bbb R}|\psi|^2dx=\Bigl|{\rm Im}\,\Bigl(\frac{k}{\sin k}\sum_{n=-\infty}^\infty f(n)\overline\psi(n)\Bigr)\Bigr|\leq  \Bigl|\frac{k}{\sin k}\Bigr| \|f\| \|u\|.
\]
Let us  show now  that there is a continuous   function $c(k)>0$ defined  on  ${\Bbb C}\setminus \pi{\Bbb Z}$
for  which
\[
\|\psi\|_{L^2}\geq   \Bigl|\frac{c(k)}{\sin k}\Bigr| \, \|u\|_{\ell^2}.
\]
Indeed,  this follows  from the fact  that
\begin{equation}\label{psi=u}
\psi(x)=\frac{\sin(k(x-n))}{\sin k}u_{n+1}-\frac{\sin(k(x-n-1))}{\sin k}u_n, \qquad \forall x\in [n,n+1],
\end{equation}
implying  the  inequality
\[
|\sin k|^2 \int_n^{n+1}|\psi|^2dx\geq c^2(k)(|u_n|^2+|u_{n+1}|^2),
\]
where 
\[
c^2(k)=\min_{0\leq \theta\leq 2\pi} \int_0^1\Bigl|\sin(kx)\sin \theta-\sin(k(x-1))\cos \theta\Bigr|^2dx
\]
  is  a  positive  continuous  function of $k$.
If $c(k)$  was equal to $0$, then one would find $\theta_0\in [0,2\pi]$ for which
\[
 \int_0^1\Bigl|\sin(kx)\sin \theta_0-\sin(k(x-1))\cos \theta_0\Bigr|^2dx=0.
\]
It is   easy to see that the latter is impossible  for any $\theta_0\in [0,2\pi]$. It remains to set  $C(k)=|k|/c(k)$.
$\,\,\,\,\Box$

\bigskip

It follows  from \eqref{normJk}  that
\begin{equation}\label{normJ}
\|J^{-1}\|\leq \frac{C_0(k)}{| {\rm Im}\, k^2|},\qquad \forall\,  |{\rm Im}\, k^2|>0,
\end{equation} where $C_0(k)>0$ is  continuous on ${\Bbb C}\setminus \pi{\Bbb Z}$.  While  it is not clear  whether $C(k)$ in \eqref{normJk} can be written explicitly,  
there is an explicit  expression for   the function $C_0(k)$  in  \eqref{normJ}.
 To show  this, we    estimate
for the norm of the operator \begin{equation}\label{R(k)}R(k)=\frac{\sin k}{k}J^{-1}.\end{equation}   

\begin{proposition}
Let $R(k)$ be the operator  defined in \eqref{R(k)}. Then  there is a universal  positive  constant
$C>0$  for which
\begin{equation}\label{boundR(k)}
\|R(k)\|\leq C\frac{(1+|{\rm Im}\, k|)}{|k||{\rm Im}\, k|}.
\end{equation}
\end{proposition}

{\it Proof}.
Since the matrix elements of the operator $R(k)$  are the numbers 
$$
\varkappa(n,m)=e^{ik|n-m|}/{2ik},
$$
we conclude that 
\[
\|R(k)\|\leq \sum_{m=-\infty}^\infty |\varkappa(n,m)| \leq 
|k|^{-1}\Bigl(1+\int_1^\infty  e^{-{\rm Im}\, k (x-1)}dx\Bigr).
\]
This leads to  the  bound \eqref{boundR(k)}.
$\Box$

\bigskip

 It  is  known that the perturbation determinant can be used to  analyze the changes in the spectral measure caused by the perturbation.
One  way to  analyze them is to  use the relation of the perturbation determinant to the spectral shift  function.  However,  since we  are interested in the absolutely continuous  spectrum, we will
use a different  property of the function $L(k)$:
namely,  that the values the perturbation  determinant at  real points $k$  are  related to the derivative of the measure $\mu_{\rm ac}$.
This relation was made  known to the broader  audience through  the paper \cite{DK} and  is  used as  a standard tool in the study of absolutely continuous spectra  of Schr\"odinger operators.

Let  $u$ be  the Jost solution of 
\eqref{1.29}.
Then $u$ is a linear  combination of two  solutions $e^{ik n}$
and $e^{-ik n}$ for $n\leq 0$:
$$
u_n=a(k)e^{ik n}+b(k)e^{-ik n}\qquad  \text{for}\quad n\leq 0.
$$
The latter  is equivalent   to the relation
$$
\psi(x)=a(k)e^{ik x}+b(k)e^{-ik x}\qquad  \text{for}\quad x\leq 0
$$
involving  the solution of \eqref{1.1}.   The coefficients $a(k)$ and $b(k)$ with $k\in {\Bbb R}$ are  often called  the scattering coefficients  because of their relation to 
the   scattering matrix
\[
S(k)=   \left( \begin{array}{cc} 
\frac{1}{a(k)} & -\frac{\overline{ b(k)}}{a(k)}\\
\frac{b(k)}{a(k)} & -\frac{1}{a(k)}
\end{array}\right).
\]
On the other  hand,   the Birman-Krein  formula \cite{BK} says that  the determinant of the scattering matrix $S(k)$
can be  expressed  in terms  of the  spectral shift  function $\xi={\rm Arg} L(k)$.
Namely,   ${\rm det}\, S(k)=(1-|b(k)|^2)/a^2(k)=e^{-2i \xi}$.
Therefore,
\[
{\rm Arg } \bigl(L(k)\bigr)={\rm Arg } \bigl(a(k)\bigr),\qquad \text{for   real}\quad k=\overline k\neq 0.
\]
So  the function $L(k)/a(k)$  is  real  on ${\Bbb R}\setminus \{0\}.$ Combining  this  fact  with  the  properties
\[
L(-k)=\overline{L(k)}\qquad \text{and}\quad a(-k)=\overline{a(k)},
\]
we  conclude that $L(-k)/a(-k)=L(k)/a(k)$.  This  tells  us   that  the function
\[
\zeta(z)=L(\sqrt z)/a(\sqrt z), \qquad  z\in {\Bbb C}\setminus \{0\}
\]
  is    analytic  on  $ {\Bbb C}\setminus \{0\}$.  To show  that it is  also analytic  at $z=0$, we    will establish the  two properties:
\begin{equation}\label{zto1}
\lim_{z\to\infty}\zeta(z)=1,
\end{equation}
while
\begin{equation}\label{xzto0}
\lim_{x\to0}x\zeta(x)=0,\qquad \text{when}\quad x\to0\quad \text{along  the positive half-line}\quad {\Bbb R}_+.
\end{equation}
The first property excludes  an essential singularity at $z=\infty$,    and hence,  at $z=0$.
The second  property tells  us that $z=0$ is not a pole. Thus, if   both \eqref{zto1} and \eqref{xzto0}  are true,  then
$\zeta$ in   analytic  on all of ${\Bbb C}$,  and  hence,  by Liouville's theorem,  it equals $1$  due to the condition \eqref{zto1}.
In other words,
\[
\eqref{zto1}\quad \text{and}\quad \eqref{xzto0}\implies \quad L(k)=a(k),\qquad \forall k\in {\Bbb C}_+.
\]

\begin{proposition}
Let ${\rm Tr}\, V\neq 0$. Then $\zeta$  possesses the properties \eqref{zto1}  and \eqref{xzto0}.
In particular,
\begin{equation}\label{a=L}
a(k)=L(k),\quad \text{for all}\quad k\in {\Bbb C}_+.
\end{equation}
\end{proposition}

{\it Proof  of \eqref{zto1}}.  Let $v_n=e^{-ikn}u_n$. Then one can  write the  equation \eqref{3.3} as 
\[
v_n=1+\frac{1}{2ik}\sum_{m=n}^\infty(1-e^{-2ik(n-m)})V(m)v_m.
\]
If $V$ has a finite support,  the solution of this equation   convereges  uniformly to $1$
as $k\to\infty$. In particular,
\[
a(k)+b(k) e^{-2ikn}\to 1,\qquad \text{as}\qquad k\to \infty\quad \text{uniformly in}\,\,n,
\]
which imlies that
$
a(k)\to 1$ and $ b(k)\to 0.
$

 To  show that 
\begin{equation}
\label{Lto1}L(k)\to 1\qquad \text{ as }k\to\infty,
\end{equation}
we represent the  function $L(k)={\rm det}\, (J_kJ^{-1})$ 
in the form
\[
L(k)={\rm det}\, \Bigl(I+\frac{\sin k}{k} VJ^{-1}\Bigr)={\rm det}\, \Bigl(I+\frac{\sin k}{k} \Omega WJ^{-1}W\Bigr),
\]
where $W=\sqrt{|V|}$ and $\Omega=V/W$  is the   sign of $V$.
This representation  implies  that, if the  Hilbert-Schmidt  norm of $\frac{\sin k}{k} \Omega WJ^{-1}W$ satisfies
\begin{equation}\label{HS<1}
\|\frac{\sin k}{k} \Omega WJ^{-1}W\|_{{\frak S}_2}<1,
\end{equation}
then
\begin{equation}\label{razlozhenie}
\log L(k)=\sum_{n=1}^\infty \frac{(-1)^{n+1}}{n}{\rm Tr}\,\Bigl(\frac{\sin k}{k} \Omega WJ^{-1}W\Bigr)^n.
\end{equation}
Since the matrix  elements of the operator $J^{-1}$
are the numbers
$
\frac1{2i\sin k }e^{ik |n-m|},
$
the first  term on the right hand  side of \eqref{razlozhenie}  equals $\frac1{2ik }{\rm Tr }\, V$.  Therefore,
\begin{equation}\label{razlozhenie2}
\log L(k)=\frac1{2ik }{\rm Tr }\, V+\sum_{n=2}^\infty \frac{(-1)^{n+1}}{n}{\rm Tr}\,\Bigl(\frac{\sin k}{k} \Omega WJ^{-1}W\Bigr)^n.
\end{equation}
Moreover,
\[
\|\frac{\sin k}{k} \Omega WJ^{-1}W\|^2_{{\frak S}_2}=\frac{1}{|2k|^2}\sum_{n,m=1}^\infty |W(n)|^2|e^{ik|n-m|}|^2 |W(m)|^2\leq \frac{1}{|2k|^2}\bigl({\rm Tr}\, |V|\bigr)^2.
\]
Consequently,  \eqref{HS<1}  is fulfilled  as long as
\[
|k|>\frac{1}{2}{\rm Tr}\, |V|,
\]
and  for such $k$'s,
\[
|\log L(k)|\leq \frac1{2|k| }{\rm Tr }\, |V|+\sum_{n=2}^\infty \frac{1}{n|2k|^n} \bigl({\rm Tr}\, |V|\bigr)^n=\log\Bigl(1- \frac1{2|k| }{\rm Tr }\, |V|\Bigr)^{-1}.
\]
The relation \eqref{Lto1}  follows. $\,\,\,\,\,\,\,\,\,\Box$

\bigskip

{\it Proof of \eqref{xzto0}}. Let $k=\bar k\neq 0$ be real, and let $u_n$  be the Jost solution of \eqref{3.3}.
Define the Wronskian
\[
W_n=u_n\overline{u}_{n-1}-u_{n-1}\overline{u}_n,\qquad n\in {\Bbb Z}.
\]
Then $W_{n+1}-W_n=0$ for all $n$, which means that $W_n$ is  constant.  The value of this  constant can be computed in two different  ways.
For large values of $n>n_0$,  we  have $V(n)=0$,  $u_n=e^{ikn }$,  and
\[
W_n=e^{ik}-e^{-ik}=2i \sin k.
\]
For  negative  values of $n<0$,  we have $u_n=a(k)e^{ikn}+b(k)e^{-ikn}$. Therefore, 
\[
W_n=2i \sin k(|a(k)|^2-|b(k)|^2).
\]
Consequently, $|a(k)|^2-|b(k)|^2=1$  for $k\in {\Bbb R}\setminus \{0\}$,  and
\[
|\zeta(k^2)|=\frac{|L(k)|}{|a(k)|}\leq |L(k)|,\qquad  \forall k=\overline k\neq 0.
\]
Thus,    \eqref{xzto0} will  be established,
once we show  that $|L(k)|=O(1/|k|)$   as $|k|\to0$. This follows  from the fact  that
\[
\frac{\sin k}{k} \Omega WJ^{-1}W=T_1(k)+T_2(k),\qquad \text{
where 
}\qquad 
T_2(k)=\frac1{2ik}\Omega w\langle \cdot, w  \rangle
\]
is the  rank $1$ operator  constructed for  the vector $w\in \ell^2({\Bbb Z})$ such that $w(n)=W(n)$ for all $n\in {\Bbb Z}$.
Writing $T_2(k)$ separately  allows one to understand  the singularity of  $\frac{\sin k}{k} \Omega WJ^{-1}W$  at zero: it follows  from the  explicit  expression for the  matrix elements  of $T_1(k)$
\[
\frac{1}{2ik} \Omega(n) W(n)(e^{ik|n-m|}-1)W(m)
\]
that the function
$T_1(k)$  is analytic on all of  ${\Bbb C}$.    Representing   $L(k)$ as  the product
\[
L(k)={\rm \det}\, (I+T_2(k))\, {\rm \det}\, \Bigl(I+(I+T_2(k))^{-1}T_1(k)\Bigr),
\]
where  ${\rm \det}\, (I+T_2(k))=1+\frac1{2ik}{\rm Tr}\, V$, we  see that
we only need to show that the function
$k\mapsto {\rm \det}\, (I+(I+T_2(k))^{-1}T_1(k))$  is  an analytic  near $k=0$.
It is easy to check that
\[
(I+(I+T_2(k))^{-1}=I-\frac1{2ik+{\rm Tr }\, V}\, \Omega w\langle \cdot, w  \rangle.
\]
It remains to note that   this  function  is analytic at $k=0$ as long  as ${\rm Tr }\, V\neq 0$. $\,\,\,\,\,\,\,\,\,\,\Box$

\bigskip
In the proposition below,  we indicate that $a$ and $L$  also   depend on  $V$ by writing
\[
a(k)=a(V,k)\qquad\text{and}\qquad L(k)=L(V,k).
\]

\begin{proposition}\label{PrVntoV}   Let ${\rm Im }\, k>0$.
Let $V_n$   be a sequence of real valued  functions on ${\Bbb N}$  that  converges to $V$ in $\ell^1({\Bbb N})$.
Let $a(V_n,k)$  and $L(V_n,k)$  be the scattering  coeficient and the  perturbation  determinant corresponding 
to  the  potential $V_n$.
Then
\[
|a(V_n,k)-a(V,k)|+|L(V_n,k)-L(V,k)|\to 0,\qquad \text{as}\quad n\to\infty.
\]
\end{proposition}

{\it Proof}. According to  \eqref{2.2}
\[
|L(V,k) -L(V_n,k))| \leq C_k\|V_n-V\|_1 \exp (C_k(\|V_n\|_1 + \|V\|_1 )+1),
\] where $C_k=\bigl|\frac{\sin k}{k}\bigr| \|J^{-1}\|$.
This  proves  that $|L(V,k) -L(V_n,k))|\to0$.

To prove that $|a(V,k) -a(V_n,k))|\to0$, we write   the equation \eqref{3.3} in a different  form.  Namely,
setting  $v_n=e^{-ikn}u_n$,  where $u_n$ is  the Jost  solution of \eqref{3.3}, we obtain
\[
v_n=1+\frac{1}{2ik}\sum_{m=n}^\infty(1-e^{-2ik(n-m)})V(m)v_m.
\]
The latter  relation  is  an equation
of the form
\[
v={\bf 1}+ T(V)\, v,\qquad v=\{v_n\}\in \ell^\infty,
\]  
where $T(V)$ is the operator on $\ell^{\infty}({\Bbb Z})$ with the   matrix  elements  
\[
T_{n,m}(V)=\begin{cases}\frac{1}{2ik}(1-e^{-2ik(n-m)})V(m),\qquad \text{if}\qquad  m\geq n\\
0,\qquad \text{if}\qquad  
m<n.
\end{cases}
\]
Note that $T(V)$ is invertible,  and $\|T(V)-T(V_n)\|\leq \frac{1}{2|k|}\|V-V_n\|_1$. Moreover,
\[
(I-T(V_n))^{-1}=\Bigl(I+(I-T(V))^{-1}(T(V_n)-T(V))\Bigr)^{-1}(I-T(V))^{-1}.
\]
Consequently,   $(I-T(V_n))^{-1}$  converges to $(I-T(V))^{-1}$,  which implies that
\[a(V_n,k)+b(V_n,k)e^{-2ikm} \to a(V,k)+b(V,k)e^{-2ikm}\quad \text{ uniformly in } m<0\]
Since ${\rm Im}\, k>0$,  one can drop   the  $b(V,k)$-terms,   because  $\lim_{m\to-\infty}e^{-2ikm}=0$.
Thus,
\[a(V_n,k)\to a(V,k),\quad \text{ as } n\to\infty.\]
$\,\,\,\,\Box$

\bigskip

As a   consequence, we  obtain  the following important result.

\begin{theorem}
Let $V:{\Bbb N}\to {\Bbb R}$ have  a finite support. Then
\begin{equation}\label{a=det}
a(k)=L(k),\qquad \text{ for  all }\quad k\in {\Bbb C}_+.
\end{equation}
\end{theorem}

{\it Proof}.  Relation \eqref{a=det}  has been already established  for  the case ${\rm Tr}\, V\neq 0$.
If ${\rm Tr}\, V=0$,   one can find a sequence $V_n$ with ${\rm Tr}\, V_n\neq 0$  that converges to $V$ in $\ell^1$.
Since
\[
a(V_n,k)=L(V_n,k),\qquad \text{ for  all }\quad k\in {\Bbb C}_+,
\]
the statement of the  theorem follows from Proposition~\ref{PrVntoV}  by taking  the limit as $n\to\infty$  on both sides.

\smallskip
$\Box$

\bigskip

The conformal map $k\mapsto k^2$ suggests replacing $m$ by 
\begin{equation} \label{1.15a}
M_\mu (k) = m(k^2).
\end{equation}
Clearly, $M_\mu$ is meromorphic on ${\Bbb C}_+=\{z:\,\,{\rm Im } \,z>0\}$ with poles at 
the points $k_j$ where 
\begin{equation} \label{1.16}
 E_j =k_j^{2}<0.
\end{equation}
If $V$ is a finite rank operator,  $M_\mu$ has boundary values everywhere on ${\Bbb R}\setminus  \pi{\Bbb Z}$,
\begin{equation} \label{1.18}
M_\mu (k)=\lim_{\epsilon\downarrow 0}\, M_\mu (k+i\epsilon )
\end{equation}
with $M_\mu (k) = \overline{M_\mu (-k)}$ and ${\rm Im}\, M_\mu (k)\geq 0$
for $k>0$, $k\notin  \pi{\Bbb Z}$.

From the integral representation \eqref{1.2},
\begin{equation} \label{1.21}
{\rm Im}\, m (E+i0) = \pi\frac{d\mu_{{\rm ac}}}{dE}
\end{equation}
so  condition \eqref{1.4} becomes
$$
\int_a^b\log[{\rm Im}\, M_\mu (k)]\, dk >-\infty
$$
for any  $0<a<b<\infty .$
Moreover, we have by \eqref{1.21} that
\begin{equation} \label{1.22}
\tfrac{2}{\pi}\int_a^b {\rm Im}\,[M_\mu (k)] k\,dk = 
	\mu_{{\rm ac}} (a^2,b^2).
\end{equation}

The  following theorem allows  us to link $|u_0|$ and $|L|$ on ${\Bbb R}$ to ${\rm Im}\,M$.

\begin{theorem} \lb{T2.17} Let $V$ be a  finite rank operator. Then for all real $k=\overline k\notin \pi{\Bbb Z}$,
\begin{equation} \label{2.68}
|u_0 |^2 {\rm Im}\, M_\mu(k)=k.
\end{equation}
Moreover {\rm ,}
\begin{equation} \label{2.69}
4|L(k)|^2  \geq \frac{ k }{{\rm Im}\, M_\mu(k)},\qquad k=\overline k\notin \pi{\Bbb Z}.
\end{equation}
\end{theorem}

{\it Proof}.
Indeed,  let $k\notin \pi {\Bbb Z}$ and  let $\psi$ be the solution of \eqref{1.1}  equal to $e^{ikx}$ to the right of the  support of $V$.
Then
\[
\psi(x)=\frac{\sin(kx)}{\sin k}u_1-\frac{\sin(k(x-1))}{\sin k}u_0,
\]
and
\[
\psi'(0)=\frac{k}{\sin k}u_1-\frac{k \cos k}{\sin k}u_0.
\]
Consequently,  
\begin{equation}\label{M=u1}
M_\mu(k)=\frac{k}{\sin k}\frac{u_1}{u_0}-\frac{k \cos k}{\sin k},\qquad \text{and}\quad {\rm Im}\, M_\mu(k)=\frac{k}{\sin k}{\rm Im}\,\Bigl(\frac{u_1}{u_0}\Bigr).
\end{equation}
On the other hand,  since $u_1\overline u_0-u_0\overline u_1=2i\sin k$,  it is easy to see that ${\rm Im}\,(\frac{u_1}{u_0})=\frac{\sin k}{|u_0|^2}$.
That proves \eqref{2.68} for $k\notin \pi {\Bbb Z}$.  

The inequality \eqref{2.69}  is a consequence of  the relation $u_0=a(k)+b(k)$  and  the bound $|b(k)|\leq |a(k)|$.
$\qquad\Box$

\bigskip

Further  arguments in the proof are  based  on the analysis  of  the terms in  the expansion of 
$\log |L(V,k)|$ into the logarithmic series.  Since odd    terms in this expansion switch their sign,  when  one changes $V$ to $-V$,
they are not present in the sum  $\log |L(-V,k)|+\log |L(V,k)|$  due to their cancellation.   
For the sake of convenience, we introduce  the notation
\[
L_4(V,k)={\rm det}_4\,\big[J_kJ^{-1}\bigr].
\]

\begin{lemma} \lb{L2.12} Let $V$ have   a  finite  support,  and let $\chi$ be   the   characteristic function of the support of $V$.  
Then 
\begin{equation} \label{2.54}\begin{split}
\log\big[L(V,k)\big]+\log\big[L(-V,k)\big] = - {\rm Tr}\, \Bigl(\frac{\sin k}{k} VJ^{-1}\chi\Bigr)^2\\
+\log\,\big[L_4(V,k)\big]+\log\,\big[L_4(-V,k)\big].
\end{split}
\end{equation}
If $k=\overline{ k}\neq 0$ is real,  then
\begin{equation}\label{Fourier}
{\rm Re}\,\Bigl( {\rm Tr}\, \bigl(\frac{\sin k}{k} VJ^{-1}\chi\bigr)^2\Bigr)=-\frac{|\hat V(2k)|^2}{4k^2},
\end{equation}
where $\hat V$ is the sum of the Fourier series
\[
\hat V(2k)=\sum_{n=1}^\infty e^{2ik n}V(n).
\]
\end{lemma}

{\it Proof}.  Indeed,   the equality
\[\begin{split}
\log\big[L(V,k)\big] = {\rm Tr}\, \Bigl(\frac{\sin k}{k} VJ^{-1}\chi\Bigr)- \frac12{\rm Tr}\, \Bigl(\frac{\sin k}{k} VJ^{-1}\chi\Bigr)^2\\
+\frac13 {\rm Tr}\, \Bigl(\frac{\sin k}{k} VJ^{-1}\chi\Bigr)^3+\log\,\big[L_4(V,k)\big]
\end{split}
\]
implies \eqref{2.54}.

To   prove \eqref{Fourier},   we   compute the following  trace  explicitely
\[
 {\rm Tr}\, \bigl(\frac{\sin k}{k} VJ^{-1}\chi\bigr)^2=\frac{-1}{4k^2}\sum_{n,m=1}^\infty V(n) e^{ik|n-m|}V(m) e^{ik|m-n|}.
\]
As a  result,  we  obtain
\[\begin{split}
{\rm Re}\,\Bigl( {\rm Tr}\, \bigl(\frac{\sin k}{k} VJ^{-1}\chi\bigr)^2\Bigr)=\frac{-1}{4k^2}\sum_{n,m=1}^\infty V(n) \cos(2k|n-m|)V(m)=\\
\frac{-1}{4k^2}\Bigl(\bigl(\sum_{n=1}^\infty V(n) \cos(2kn)\bigr)^2-\bigl(\sum_{n=1}^\infty V(n) \sin(2kn) \bigr)^2\Bigr),
\end{split}
\]
which coincides  with the right hand  side of \eqref{Fourier}  because $V=\overline V$ is real.
$\qquad \Box$

\bigskip

For  the sake of  completeness, we prove the  following statement.

\begin{lemma} \lb{L2.11} Let $V$ generate  a  finite  rank operator on $\ell^2({\Bbb N})$. 
That is, there is a number $n_0\in {\Bbb N}$ such that  $V(n)=0$ for all $n>n_0$. Then for  any $\varepsilon>0$,
\begin{equation} \label{2.52}
\log\,{\rm det}\,\big[J_kJ^{-1}\big] \sim \sum_{n=1}^\infty \frac{(-1)^{n+1}}{n\, (2ik)^n} {\rm Tr}\, V^n,
\end{equation}
as $k\to\infty$  inside the sector
\begin{equation} \label{2.53}
\{k\mid \,\, \varepsilon<{\rm Arg}\, k<\pi-\varepsilon\}.
\end{equation}
\end{lemma}

\medskip

{\it Proof.} Note that  for large  values of $k$  that belong  to  the sector \eqref{2.53},   the trace norm of the operator $\frac{\sin k}{k} VJ^{-1} $  obeys $$\|\frac{\sin k}{k} VJ^{-1} \|_{{\frak S}_1}<1.$$
This follows  from the fact  that
\[
J^{-1}\sim e^{ik} I+O(e^{2ik}),\qquad \text{as}\quad k\to \infty.
\]
Consequently,  for such $k$'s,
\[
\log\,{\rm det}\,\big[J_kJ^{-1}\big] = \sum_{n=1}^\infty \frac{(-1)^{n+1}}{n} {\rm Tr}\, \Bigl(\frac{\sin k}{k} VJ^{-1}\Bigr)^n.
\]
It remains to note that if $k$  stays inside the set \eqref{2.53},   then
\[
{\rm Tr}\, \Bigl(\frac{\sin k}{k} VJ^{-1}\Bigr)^n={\rm Tr}\, \Bigl(\frac{1}{2ik} V\Bigr)^n+O(e^{ik}),\qquad \text{as}\quad k\to \infty.
\]
$\Box$

\bigskip

Lemmas~\ref{L2.12} and ~\ref{L2.11} show  that  the behavior of $\log L(V,k)$ might change depending on whether the point $k$ moves  along the real  line  or the imaginary line.
 It decays along the imaginary direction, but does not  necessarely decay along the real axis.
  In these circumstances,  obtaining   standard trace formulas (containing intergals  of $\log|L(k)|$ over  the whole real  line)   becomes less  interesting.  
Instead,  we  will  obtain inequalities   in which  $\log|L(k)|$  is integrated  over  a bounded interval.

For $0<\alpha<\beta<\infty$, define  the polynomial $p(k)$ by
\begin{equation}\label{defp}
p(k)=(k-\alpha)^5(\beta -k)^5, \qquad k\in {\Bbb C}.
\end{equation}
Observe  that $p(k)\geq 0$  on $[\alpha,\beta]$.
We will shortly need  the following result.
\begin{proposition}
Let $V$ be  of  finite rank. Then

{\rm (i) } The  function
\[
L_4(V,k)={\rm det}_4\bigl( J_kJ^{-1}\bigr)
\]
extends  analytically into the region ${\Bbb C}\setminus \{0\}$.

{\rm (ii) }  The zeros  of $L_4(V,k)$   coinside with  the zeros of $L(k)$ and are  imaginary.

{\rm (iii)} For any $0<\alpha<\beta<\infty$  having the property  $\alpha,\beta\notin \pi{\Bbb N}$,
\begin{equation}\lb{intV4}
\Bigl| \int_{\alpha}^\beta\log\bigl[L_4(V,k)\bigr]p(k) dk\Bigr|\leq C\|V\|^4_4.
\end{equation}
with a  positive constant $C$ depending only on $\alpha$  and $\beta$.
\end{proposition}

{\it Proof.} {\rm (i) } 
 Recall that
\[
\log\,{\rm det}_4\,\big[J_kJ^{-1}\big]=\log\,{\rm det}_4\,\big[I+VR(k)\big]=\log\,{\rm det}_4\,\big[I+VR(k)\chi\big],
\]
where $R(k)=\frac{\sin k}{k}J^{-1}$   and $\chi$  is the characteristic  function of the ``support'' of $V$, that is   the set of integers  at  which $V\neq 0$.
Since $VR(k)\chi$ can be extended  analytically  into ${\Bbb C}\setminus \{0\}$,  so can  $L_4(V,k)$.

Assertion {\rm (ii) }  is a consequence  of a  very well known fact that holds  for  any  $T\in {\frak S}_1$:
\[
{\rm det}\,(I+T)=0\qquad  \Longleftrightarrow \qquad {\rm det}_4\,(I+T)=0.
\]

{\rm (iii) } 
While
 involvement  of the  parameter $t$ in the  proof  is not  obvious,
we still define
\[
\Delta(t)=\log\,{\rm det}_4\,\big[I+t\, VR(k)\big].
\]
We intend to  use  the simple estimate:
\begin{equation}\label{det<1/8}
|\Delta(1)|=|\Delta(1)-\Delta(0)|\leq \int_0^1 |\Delta'(t)|dt.
\end{equation}
A  straightforward  computation shows  that
\[\begin{split}
\Delta'(t)=
\frac{d}{dt}\Bigl( \log\,{\rm det}_4\,\big[I+t\, VR(k)\big]\Bigr)={\rm Tr}\, \Bigl(\bigl(I+t\, VR(k)\bigr)^{-1}V R(k)\Bigr)-\qquad 
\\
\sum_{n=1}^3(-t)^{n-1}{\rm Tr}\, \bigl(V R(k)\bigr)^n=t^3\,{\rm Tr}\, \Bigl(\bigl(I+t\,VR(k)\bigr)^{-1}(V R(k))^4 \Bigr)\quad \\
=t^3\,{\rm Tr}\, \Bigl( J\bigl(J+\frac{\sin k}{k} t\, V\bigr)^{-1}(V R(k))^4\Bigr).\qquad \qquad  \qquad
\end{split}
\]
Thus,
\[
\Bigl| \frac{d}{dt}\Bigl( \log\,{\rm det}_4\,\big[I+t\, VR(k)\big]\Bigr) \Bigr|\leq   \|J\|  \, \| \bigl(J+\frac{\sin k}{k} t\, V\bigr)^{-1}\| \, \| R(k)\|^4 \|V \|^4_4  .
\]
Note  now that $\bigl(J+\frac{\sin k}{k} t\, V\bigr)^{-1}$  is the   operator $J_k^{-1}$  with $V$ replaced  by $t\, V$.
Consequently,   \eqref{normJk} holds  with $J_k^{-1}$ replaced  by the   operator  $\bigl(J+\frac{\sin k}{k} t\, V\bigr)^{-1}$  as  long as $t$ stays  real.
Taking  into account   the inequalities  \eqref{normJk} and \eqref{boundR(k)},   we obtain
\begin{equation}\label{det<}
\Bigl| \frac{d}{dt}\Delta(t) \Bigr|\leq    \frac{C(k)\, |t|^3}{|{\rm  Re}\, k||{\rm  Im}\, k|^5} \|V \|^4_4,  \qquad \qquad  \qquad
\end{equation}
where $C(k)>0$ is a  continuous   function of $k$ on the set ${\Bbb C}\setminus \pi{\Bbb Z}$.
Integrating \eqref{det<}  from $0$ to $1$ and taking into account \eqref{det<1/8}, we    conclude
that 
\begin{equation}\label{det<V4}
\Bigl|\log\,{\rm det}_4\,\big[J_kJ^{-1}\big]\Bigr|\leq   \frac{C(k)\|V\|^4_4}{4\, |{\rm  Re}\, k||{\rm  Im}\, k|^5} \qquad \text{if}\quad \alpha<{\rm Re}\, k<\beta.
\end{equation}
Since the   function $\log L_4(V,k)$ is analytic  inside a region that   does not  intersect   the   imaginary  axis,
the value of the integral  of  this function over a contour  contained in such a  domain  equals zero.
Therefore,
\begin{equation}\label{contour}
\int_{\alpha}^\beta\log\bigl[L_4(V,k)\bigr]p(k) dk=\int_{C_{\alpha,\beta}}\log\bigl[L_4(V,k)\bigr]p(k) dk,
\end{equation}
where $C_{\alpha,\beta}$ is the  half-circle $\{k\in {\Bbb C}:\,\, |k-\frac{(\alpha+\beta)}2|=\frac{(\beta-\alpha)}2, \,\,{\rm Im}\,k\geq0 \}$
connecting the two points $\alpha$ and $\beta$.

The   statement (iii) follows  now  from  \eqref{defp}, \eqref{det<V4} and \eqref{contour}.
$\Box$

\bigskip

\bigskip

\begin{theorem}
Let $V$  be a finite rank operator.   Let $0<\alpha<\beta<\infty$ be two points having the property $\alpha,\beta\notin \pi {\Bbb Z}$
and  let $p(k)$ be   the function defined  by \eqref{defp}.
Then
\begin{equation}\lb{traceineq}
\int_{\alpha}^\beta \log\bigl[\frac{k}{4\,{\rm Im}\, M_\mu(k)}\bigr]\,p(k)dk\leq \int_{\alpha}^\beta \frac{|\hat V(2k)|^2}{k^2}\,p(k)dk+C\|V\|^4_4
\end{equation}
with a constant $C>0$  that depends only on $\alpha$ and $\beta$.
\end{theorem}

{\it Proof}.
Since $|L(V,k)|\geq 1$,  we  infer  from \eqref{2.69}, \eqref{2.54} and \eqref{Fourier}  that for any $0<\alpha<\beta<\infty$ having the property $\alpha,\beta\notin \pi {\Bbb Z}$,
\[\begin{split}
\int_{\alpha}^\beta \log\bigl[\frac{k}{4\,{\rm Im}\, M_\mu(k)}\bigr]\,p(k)dk\leq 2 \int_{\alpha}^\beta \log|L(V,k)|\,p(k)dk\leq \qquad \qquad \\
2 \int_{\alpha}^\beta \log|L(V,k)|\,p(k)dk+2 \int_{\alpha}^\beta \log|L(-V,k)|\,p(k)dk=\int_{\alpha}^\beta \frac{|\hat V(2k)|^2}{k^2}\,p(k)dk+\\
 2 \, {\rm Re}\,\int_{\alpha}^\beta\Bigl( \log\,{\rm det}_4\,\big[L_4(V,k)\big]+\log\,{\rm det}_4\,\big[L_4(-V,k)\big] \Bigr)p(k) dk.\qquad \quad
\end{split}
\]
Now \eqref{traceineq}  follows  from    \eqref{intV4}. $\,\,\Box$

\bigskip


\vglue-8pt
\section{Entropy and its upper semicontinuity }
\vglue-4pt
 
The left  hand  side 
of  the inequality \eqref{traceineq} is  an 
integral of the  logarithm:
\begin{equation} \label{5.1}
Z(\mu) =\int_\alpha^{\beta} \log \biggl( \frac{k}{4\,{\rm Im}\, M_\mu(k)}
\biggr) p(k)\, dk.
\end{equation}
Taking  into account  the fact that  $M_\mu$ is related to the original spectral measure on $\sigma (H)\supset 
[0,\infty)$ as
\begin{equation} \label{5.3}
{\rm Im}\, M_\mu(k) =  \pi\,\frac{d\mu_{{\rm ac}}}{dE}\, (k^2),
\end{equation}
one  rewrites \eqref{5.1} as
\begin{equation} \label{5.4}
Z(\mu) =  \int_{\alpha}^\beta \log \biggl( \frac{\sqrt{E}}{2\pi\, d\mu_{{\rm ac}}/dE}\biggr) p(\sqrt{E})
\frac{dE}{2\sqrt{E}}.
\end{equation}
The main goal of  this section is to 
prove that, if $\mu_n\to \mu$ weakly,  then $Z(\mu_n)$  obeys
\begin{equation} \label{5.5}
Z(\mu) \leq \liminf Z(\mu_n),
\end{equation}
that is, that $Z$  is  weakly lower semicontinuous. This will let us prove  Theorem~\ref{T1}.

The lower semicontinuity of such integrals  was deduced in \cite{KS} by providing a variational principle
that allows one  to rewrite $Z$ as   the supremum of  weakly continuous functionals.  

\bigskip

{\bf Definition}.  Let $\nu,\mu$ be finite Borel measures on a compact Hausdorff 
space, $X$. We define the entropy of $\nu$ relative to $\mu$, $S(\nu\mid \mu)$, by 
\begin{equation} \lb{5.8}
S(\nu\mid\mu) = \left\{ \begin{array}{ll} 
-\infty &\hbox{if $\nu$ is not $\mu$-ac} \\
-\int \log (\frac{d\nu}{d\mu}) d\nu &\hbox{if $\nu$ is $\mu$-ac}.
\end{array}\right. 
\end{equation}

\bigskip

If $d\nu = f\, d\mu$, then
\begin{equation} \label{5.9}
S(\nu\mid\mu) = -\int f\log(f)\, d\mu
\end{equation}
is the more usual formula for entropy.

For the sake  of completeness of the explanation,  we   copy   the following lemma   from \cite{KS}.

\begin{lemma}\lb{L5.1} Let $\nu$ be a probability measure. Then   
\begin{equation} \label{5.10}
S(\nu\mid\mu) \leq \log\mu(X).
\end{equation}
In particular{\rm ,} if $\mu$ is  also a probability measure, then
\begin{equation} \label{5.11}
S(\nu\mid\mu) \leq 0.
\end{equation}
Equality  in {\rm \eqref{5.11}} holds if and only if $\nu=\mu$.
\end{lemma}

According  to  \eqref{5.10},
the integral  in \eqref{5.8} can diverge only to $-\infty$, not to $+\infty$.

The key to understanding of  the   semicontinuity of the entropy is  the following  variational principle
(see \cite{KS}  for its  proof).

\begin{theorem} \lb{T5.2} For all measures $\nu,\mu$,
\begin{equation} \label{5.14a}
S(\nu\mid\mu) =\inf \biggl[ \int F(x)\, d\mu - \int (1+\log F)\, d\nu(x)\biggr] 
\end{equation}
where the infimum is taken over all real-alued continuous functions $F$ having the property
$\min_{x\in X} F(x)\,>0$.
\end{theorem}

As an infimum of continuous functionals is upper semicontinuous,  we   have  the  following  remarkable and useful  result   established by Killip and Simon in \cite{KS}.

\begin{theorem} \lb{C5.3} The entropy $S(\nu\mid\mu)$ is jointly weakly upper semicontinuous in 
$\nu$ and $\mu${\rm ,} that is{\rm ,} if $\nu_n \stackrel{w}{\longrightarrow} \nu$ and $\mu_n
\stackrel{w}{\longrightarrow}\mu${\rm ,} 
then 
$$
S(\nu\mid\mu) \geq \limsup_{n\to\infty} S(\nu_n\mid\mu_n).
$$
\end{theorem}

In our applications, $\nu_n=\nu$ will be a constant sequence.
To apply this to $Z$, we note

\begin{proposition} \lb{P5.4} 
   Let $\nu$ and $\tilde \mu$  be   the two measures defined on the interval $[\alpha,\beta]$ by
\begin{equation} \label{5.16x}
d\nu (E) =  \frac{p(\sqrt{E})}{2\sqrt{E}}dE\,\quad \text{and}\quad d\tilde \mu (E) =  \frac{\pi p(\sqrt{E})}{E}d\mu,
\end{equation}
where $p(k)$  is defined   by \eqref{defp}.
Then
\begin{equation} \label{5.17x}
Z(\mu) = -S(\nu\mid\tilde \mu).
\end{equation}
\end{proposition}

Given this proposition,  Lemma~\ref{L5.1}, and Theorem~\ref{C5.3}, we have 

\begin{theorem} \lb{T5.5}  For any measure $\mu$,
\begin{equation} \label{5.20}
Z(\mu) > -\infty.
\end{equation}
If $\mu_{n}\to \mu$ weakly on ${\Bbb R}$,  then
\begin{equation} \label{5.21}
Z(\mu)\leq \liminf \,Z(\mu_n).
\end{equation}
\end{theorem}

We will call \eqref{5.21}  lower semicontinuity of $Z$.

\section{Convergence of  spectral measures}

Consider  now the matrix
\begin{equation} \lb{tildeJ}
\tilde J_k= \left( \begin{array}{ccccc}
b_1 & -1 & 0 & 0 & \cdots \\
 -1& b_2& -1& 0 & \cdots \\
 0 & -1 & b_3 & -1& \cdots \\
 \vdots & \vdots & \vdots & \vdots & \ddots 
\end{array}\right)\quad
\text{with}\quad b_n=2\cos k +\frac{\sin k}{ k } V(n),\quad \forall n \in {\Bbb N}
\end{equation} 
that defines a  bounded  operator on $\ell^2({\Bbb N})$. Let $\delta_n$ be the  standard vector in $\ell^2({\Bbb N})$
whose  components are  equal to zero except for the $n$-th component equal to $1$.

\begin{proposition} Let ${\rm Im}\, k^2>0$, and
let the function $V:{\Bbb N}\to {\Bbb R}$ be  bounded.  
Then  the operator  $\tilde J_k$ has a bounded  inverse  obeying 
\begin{equation}\label{normtildeJk}
\|\tilde J_k^{-1}\|\leq C(k),\qquad \forall\,  |{\rm Im}\, k^2|>0,
\end{equation}
with a  constant $C(k)>0$  depending only on $k$.  Moreover,   if  $u_n$  is the Jost  solution,  then
\begin{equation}\label{J=u1u0}
(\tilde J_k^{-1} \delta_1,\delta_1)=\frac{u_1}{u_0}.
\end{equation}
\end{proposition}

{\it Proof}.  To establish invertibility of $\tilde J_k$   we  follow  the steps of the proof 
of Lemma~\ref{LnormJ}.  The only  difference   is that  now we  need to work with the operator on the half-line.
Namely,
consider  the equation 
\[
\tilde J_k \phi=f,\qquad  \text{for}\quad \phi,f\in \ell^2({\Bbb N}).
\]
For the sake of  convenience,  set $\phi_0=0$.
Since the values on  boundary of  $[n,n+1]$ determine the   solution of the equation   $-\psi''=k^2\psi$ inside the interval,
we can find a  continuous function $\psi\in L^2[0,\infty)$    such that 
\[
-\psi''=k^2\psi\qquad \text{  a.e.  on }\quad {\Bbb R},\quad \text{and}\quad \psi(n)=\phi_n,\qquad \forall n\in {\Bbb N}\cup \{0\}.
\]
In this case,    the  jumps of   the function $\psi'$
have the following property:
\[
\psi'(n+0)-\psi'(n-0)=V(n)\psi(n)+\frac{k}{\sin k} f(n),\qquad \forall n\in {\Bbb N}.
\]
Using this property and  integrating by parts, we obtain    that
\[
\int_0^\infty |\psi'|^2dx+\sum_{n=1}^\infty V(n)|\psi(n)|^2=k^2 \int_0^\infty|\psi|^2dx+\frac{k}{\sin k}\sum_{n=1}^\infty f(n)\overline{\psi(n)}.
\]
Consequently,
\[
\bigl|{\rm Im}\, k^2\big| \int_0^\infty|\psi|^2dx=\Bigl|{\rm Im}\,\Bigl(\frac{k}{\sin k}\sum_{n=1}^\infty f(n)\overline\psi(n)\Bigr)\Bigr|\leq  \Bigl|\frac{k}{\sin k}\Bigr| \|f\| \|\phi\|.
\]
Finally,   since 
\begin{equation}\notag
\psi(x)=\frac{\sin(k(x-n))}{\sin k}\phi_{n+1}-\frac{\sin(k(x-n-1))}{\sin k}\phi_n, \qquad \forall x\in [n,n+1],
\end{equation}
there is a constant   $c(k)>0$
for  which
\[
\|\psi\|_{L^2}\geq c(k)\, \|\phi \|_{\ell^2}.
\]
The latter follows  from the fact    established in the proof of Lemma~\ref{LnormJ}:
\[
|\sin k|^2 \int_n^{n+1}|\psi|^2dx\geq c_0(k)(|\phi_n|^2+|\phi_{n+1}|^2),
\]
where 
\[
c_0(k)=\min_{0\leq \theta\leq 2\pi} \int_0^1\Bigl|\sin(kx)\sin \theta-\sin(k(x-1))\cos \theta\Bigr|^2dx
\]
  is  a  positive  constant  depending only on $k$.

 To establish \eqref{J=u1u0},   define
\[
v=\tilde J_k^{-1} \delta_1.
\]
Since $v$ is an $\ell^2({\Bbb N})$-solution,  there is a nonzero constant $c$  for which
$v_n= cu_n$.  On the other hand,  $c$ must be equal to $u_0^{-1}$, because
\[
-v_2+b_1 v_1=1,\qquad \text{while}\quad
-u_2+b_1 u_1-u_0=0.
\]
Thus,
\[
v_1=(\tilde J_k^{-1} \delta_1,\delta_1)=\frac{u_1}{u_0}.
\]
$\Box$

\bigskip

\begin{corollary}
Let $\mu$  be the spectral measure of the operator $H$, and let
$m(E)$  be the function defined  by \eqref{1.2}. Then
\begin{equation}\label{m=J}
m(E)=\frac{k}{\sin k}(\tilde J_k^{-1} \delta_1,\delta_1)-\frac{k \cos k}{\sin k},
\end{equation}  with  $k^2=E\notin {\rm supp}\,\mu$.
\end{corollary}

{\it Proof}.  For ${\rm Im}\, E>0$,  formula \eqref{m=J} follows  from \eqref{M=u1} and \eqref{J=u1u0}.
It  extends to   the  general case  by analyticity.
$\Box$

\bigskip

For $n\in {\Bbb N}$,
let $V_n$ be the  truncation of $V$ given by 
\[
V_n(j)=\begin{cases}
V(j)\qquad \text{for}\quad j\leq n\\
\,0\qquad \quad\, \text{ for}\quad j> n.
\end{cases}
\] 
Define the operator $H_n$ as  the operator $H$  with $V$ replaced  by $V_n$, that is,
\begin{equation}\label{Hn}
H_n=-\frac{d^2}{dx^2}+\sum_{j=1}^nV(j)\delta(x-j).
\end{equation}

\begin{lemma} Let $V\in \ell^4({\Bbb N})$. Then
the  spectral measures  $\mu_n$  of the operators \eqref{Hn}  converge  weakly to $\mu$,   the spectral measure of $H$.
\end{lemma}

{\it Proof}.   Let $m_n(E)$  be  the  function \eqref{1.2} with $\mu$ replaced  by $\mu_n$.
 Let  also  $\tilde J_{k,n}$ be  the operator $\tilde J_k$  with $V_n$ instead of  $V$.
As $\tilde J_{k,n}\to \tilde J_k$ in ${\frak S}_4,$ we     conclude that $\tilde J_{k,n}^{-1}$ converges (in norm) to
$\tilde J_{k}^{-1}$ for all $k^2\in{\Bbb C}\setminus{\Bbb R}$. 
We infer  from the   formula \eqref{m=J}  that the  sequence of  functions $m_n$
converges to $m$  uniformly on compact sets in the upper half-plane.
This implies that $\mu_n\to\mu$ weakly. $\,\,\,\,\Box$

\section{Finite support approximations of  the potential}

Here,  we   construct  a  special sequence of  approximations of  the potential $V$.
These     approximations $V_n$ will be the  functions for which  the quantities appearing on  the right hand side of   \eqref{traceineq}
are  bounded uniformly in $n$.  It is not clear  whether 
the simple truncations of the potential $V$  have such a property.   So, the construction   requires a delicate work.

\begin{theorem}
Given $V$ satisfying the  conditions of Theorem~\ref{T1},  there is  a  sequence  $V_n$
of real valued   functions on ${\Bbb N}$ having the following properties:

{\rm (i) } The  support of each $V_n$ is  a finite subset of ${\Bbb N}$.

{\rm (ii)} The   sequence of  functions $V_n$ converges  to $V$ pointwise on ${\Bbb N}$.

{\rm (iii)} For any $[\alpha',\beta'] \subset (\alpha,\beta)$,
\begin{equation}\label{Vnapproximations}
\int_{\alpha'}^{\beta'} \bigl|\hat V_n(2k)\bigr|^2dk+\|V_n\|_4^2\leq C\Bigl( \int_{\alpha}^{\beta} \bigl|\hat V(2k)\bigr|^2dk+ \|V\|_4^2\Bigr)
\end{equation}
with a  positive universal  constant $C>0$.

\end{theorem}

{\it Proof}.  In the same way as  before,  we  extend $V$   to all of ${\Bbb Z}$ by setting $V(n)=0$ for $n\leq 0$.   Then one can  decompose  such $V$ into the sum of real-valued  functions on ${\Bbb Z}$
\[
V= V_++ W, \qquad   \text{where }\quad  V_+\in \ell^2({\Bbb Z}),\quad \text{and}\quad \hat W(2k)=0,\quad  \forall k\in [\alpha,\beta].
\]
For instance, one  can set
\[
\hat W(2k)=\begin{cases}
\hat V(2k),\quad \text{if}\quad k\notin [\alpha,\beta]\cup[-\beta,-\alpha] \quad \text{mod}\,\pi\\
\quad 0, \quad\quad  \text{ if}\quad k\in [\alpha,\beta]\cup[-\beta,-\alpha] \quad \text{mod}\,\pi,
\end{cases}
\]  where  the  equalities are understood  in the sense of distributions.  In this case,  $V_+$ has  the property $\|V_+\|^2_2= (2\pi)^{-1}\int_{\alpha}^{\beta} \bigl|\hat V(2k)\bigr|^2dk$.
Since  the restrictions of $V$ and $V_+$  to ${\Bbb N}_-:={\Bbb Z}\setminus {\Bbb N}$  are square  summable,    so is the restiction of $W$.
Moreover, recalling  that $V\bigr|_{{\Bbb N}_-}=0$,  we obtain the relation
\[
\|W\bigr|_{{\Bbb N}_-}\|_2\leq\|V_+\|_2=\Bigl((2\pi)^{-1}\int_{\alpha}^{\beta} \bigl|\hat V(2k)\bigr|^2dk\Bigr)^{1/2}.
\]
Thus, if we   replace  $W\bigr|_{{\Bbb N}_-}$  and $V_+\bigr|_{{\Bbb N}_-}$  by zero,  the condition 
\begin{equation}\label{hatW=0}
 \hat W(2k)=0,\quad  \forall k\in [\alpha,\beta]
\end{equation}
 would   change to     the inequality
\[
\int_{\alpha}^{\beta} \bigl|\hat W(2k)\bigr|^2dk\leq \int_{\alpha}^{\beta} \bigl|\hat V(2k)\bigr|^2dk.
\]

However,  we  will first consider the case \eqref{hatW=0}.  Since we only know that $W\in\ell^4$,  its  Fourier   transform  $\hat W$ is  a  distribution
that does  not have to be a function outside of   the intervals $[\alpha/2,\beta/2]$  and $[-\beta/2, -\alpha/2]$.
There is a standard  method allowing one to  turn   $\hat W$ into a continuous function.
Namely,   let   us    choose  a positive  $h\in C_0^\infty(-1,1)$ having the properties
\[
\int_{-1}^1 h(k)dk=1,\quad \text{and}\quad h(-k)=h(k)\quad \forall k\in (-1,1),
\]
and set $h_\varepsilon(k)=\varepsilon^{-1}h(k/\varepsilon)$ for $0<\varepsilon<1$.
If $\varepsilon<\frac12 \min\{|\alpha-\alpha'|,|\beta-\beta'|\}$,  then
the  support of the function
\[
\hat W_\varepsilon(k) :=\int_{-\infty}^\infty h_\varepsilon(k-k')\hat W(k')  dk'
\] does not intersect  the set $[\alpha'/2,\beta'/2]\cup[-\beta'/2,-\alpha'/2] $.
Put  differently,  $W_\varepsilon$  defined  by
\[
W_\varepsilon(n) = \hat h_\varepsilon(n) W(n),\qquad \text{with}\quad \hat h_\varepsilon(n)=\int_{-\pi}^\pi e^{-ink}h_\varepsilon(k)dk,
\]
is  a  real-valued function whose  Fourier transform  $\hat W_\varepsilon$ vanishes  on  the set $[\alpha'/2,\beta'/2]\cup[-\beta'/2,-\alpha'/2] $.
Note  that 
\[
\hat h_\varepsilon(n)\to1,\qquad \text{as}\quad \varepsilon\to0,\qquad \text{ and}\quad |\hat h_\varepsilon(n)|\leq1,\qquad \forall n\in {\Bbb Z}.
\]

To finish the proof  of the  theorem,  assume that $\delta>0$ and $N_1\in {\Bbb N}$ are given.
Then we can   choose $\varepsilon>0$ so that $|W_\varepsilon(n)-W(n)|<\delta$  for all $n\leq N_1$.
After that, we select a natural number  $N>N_1$   for which 
\[
\sum_{n=N}^\infty |W_\varepsilon(n)|^2\leq \Bigl(\sum_{n=N}^\infty |\hat h_\varepsilon(n)|^4\Bigr)^{1/2}\Bigl(\sum_{n=N}^\infty |W(n)|^4\Bigr)^{1/2}<\delta \|W\|_4^2.
\]
Define $V_{\delta,N_1}$ on ${\Bbb N}$ by
\[
V_{\delta,N_1}(n)=\begin{cases}
W_\varepsilon(n)+V_+(n),\qquad \text{ if}\quad n\leq N\\
\quad 0,\qquad \qquad\qquad \quad  \text{ if}\quad n>N.
\end{cases}
\]
Then  
\[
\|V_{\delta, N_1}||_4\leq \|W_\varepsilon\|_4+\|V_+\|_4\leq\|W\|_4+\|V_+\|_2\leq \|V\|_4+2\|V_+\|_2,
\]
which implies the  estimate
\begin{equation}\label{Vd<V4}
\|V_{\delta, N_1}||_4\leq  \|V\|_4+\sqrt{\frac2{\pi}}\Bigl( \int_{\alpha}^{\beta} \bigl|\hat V(2k)\bigr|^2dk\Bigr)^{1/2}.
\end{equation}
On the other hand,
\[
 \int_{\alpha'}^{\beta'} \bigl|\hat V_{\delta, N_1}(2k)\bigr|^2dk\leq  2\int_{\alpha'}^{\beta'} \bigl|\hat W_{\varepsilon}(2k)+\hat V_+(2k)\bigr|^2dk+4\pi\|W_{\varepsilon}+\hat V_+-V_{\delta, N_1}\|^2,
\] which leads to
\begin{equation}\label{ft<v4}
 \int_{\alpha'}^{\beta'} \bigl|\hat V_{\delta, N_1}(2k)\bigr|^2dk\leq 4\pi \|V\|_4^2+10\int_{\alpha}^{\beta} \bigl|\hat V(2k)\bigr|^2dk.
\end{equation}
We finally  define  the  sequence  of  approximations $V_n$  by setting
\[
V_n=V_{\delta, N_1},\qquad \text{with }\quad \delta=\frac1n\quad \text{and }\quad N_1=n,
\]
for each index $n\in {\Bbb N}$.  Then \eqref{Vd<V4} and \eqref{ft<v4} imply \eqref{Vnapproximations},  so all  conditions  (i),(ii), and (iii) are  fulfilled.
$\,\Box$

\section{Proofs of Theorem~\ref{T2} and Theorem~\ref{T4*}}

For a  compact operator $T$ on a  separable Hilbert space ${\frak H}$,  we denote  by $s_j(T)$  its  singular  values (s-numbers). 
In other words, 
$s_j(T)$ are eigenvalues of  the operator $\sqrt{T^*T}$  enumerated in the  decreasing order
\[
s_1(T)\geq s_2(T)\geq \dots s_j(T)\geq\dots
\]
If the sequence of  singular values  is finite,  we  extend it   by zero.

Theorem~\ref{T2} is a consequence of a  stronger  result stated  below.
\begin{theorem} \label{T3} Let $\tilde H$  be the operator defined by the quadratic form \eqref{3.1}.
Assume that 
\[
\sum_{n=1}^\infty |V(n)|<\infty.
\]
Then  the negative eigenvalues $E_j$ of the operator $\tilde H$  obey
\[
\sum_j\sqrt{|E_j|}\leq \frac12\sum_{n=1}^\infty |V(n)|.
\]
\end{theorem}

Without  any loss  of  generality,  we  may assume  that  the function $V:{\Bbb N}\to {\Bbb R}$  has a  finite  support, so  that
the corresponding multiplication by $V$ is an operator of  finite  rank on $\ell^2$. Since  $V\geq -|V|$,  and the eigenvalues  of $\tilde H$ are  monotone   functions of  $V$, 
it is enough to consider  the case   where $V\leq 0$.
Therefore, we may assume that   there is a   nonnegative  function $W$ on ${\Bbb Z}$  for which  $V=-W^2\leq 0$.
We need an appropriate version of the Birmnan-Schwinger principle suitable for perturbations considered in the paper. This version is given below:
\begin{proposition} Let $V=-W^2$.  The point
$E=-\varepsilon^2<0$  is an eigenvalue   of the operator $\tilde H$
if  and    only if $1$ is  an eigenvalue  of the operator $X_\varepsilon=WR(i\varepsilon)W$.  The multiplicities of the eigenvalues    for both operators  are  equal to $1$.
\end{proposition}

{\it Proof}.   Let $k$ be a  point in the upper half-plane,  that is,  ${\rm Im}\, k>0$.
Then  $k^2$ is an eigenvalue of $\tilde H$ if and  only if $0$ is an eigenvalue of $J_kJ^{-1}= I+VR(k)$
of the same  multiplicity.  
Since  the nonzero eigenvalues  of $-VR(k)$  and $WR(k)W$  are the same, we  obtain the   statement of this proposition.  $\,\,\Box$

\bigskip

To move  further,  we
observe that the matrix elements of the operator $X_\varepsilon$  are
\[
W(n)\frac{e^{-\varepsilon|n-m|}}{2\varepsilon}W(m)=\frac1{2\pi}W(n)\int_{\Bbb R}\frac{e^{i\xi(n-m)}d\xi}{|\xi|^2+\varepsilon^2}W(m).
\]
This relation can  be interpreted as 
\begin{equation}\label{repintY}
X_\varepsilon=\int_{\Bbb R} Y_\varepsilon(\xi)d\xi,
\end{equation}
where $Y_\varepsilon(\xi)$ is the rank one operator whose matrix elements are 
\[
\frac1{2\pi}W(n)\frac{e^{i\xi(n-m)}}{|\xi|^2+\varepsilon^2}W(m).
\] This representation is  useful   for several reasons.   First of all,
it immideately implies that
\begin{equation}\label{tr<12V}
\sum_{j=1}^\infty s_j(X_\varepsilon)\leq \int_{\Bbb R} \sum_{j=1}^\infty s_j(Y_\varepsilon(\xi))d\xi= \frac12\sum_n |V(n)|.
\end{equation}
However,  a more   important  implication of \eqref{repintY}
follows from  the group  property of  the 
function
\[
P_\varepsilon(\xi)=\frac1{\pi}\frac{\varepsilon}{|\xi|^2+\varepsilon^2}=-\frac1{\pi}{\rm Im}\,\frac{1}{\xi+i\varepsilon},\qquad \xi\in {\Bbb R}.
\]
Namely,
\[
\int_{\Bbb R} P_\varepsilon(\eta)P_\tau(\xi-\eta)\, d\eta=P_{\varepsilon+\tau}(\xi),\qquad \forall \xi\in {\Bbb R}
\]
Using this property,   we  derive the equality
\[
Y_{\varepsilon+\tau}(\xi)=\int_{\Bbb R} U^*(\xi-\eta)Y_{\varepsilon}(\eta) U(\xi-\eta) P_{\tau}(\xi-\eta)d\eta,
\]
where $U(\xi):\ell^2({\Bbb Z})\to\ell^2({\Bbb Z})$  is the unitary operator of multiplication by  the function $e^{-i\xi n}$, that is,
\[
\Bigl[U(\xi) \psi\Bigr] (n)= e^{-i\xi n} \psi(n),\qquad  \forall n \in  {\Bbb Z}.
\]
We are in a good  position  to prove the following result.  Following the  articles \cite{HLW} and \cite{HLT},   we  call  it  ``Monotonicity lemma''.
\begin{lemma}
Let $\varepsilon>0$ and $\tau>0$.
Then for each $n\in {\Bbb N}$,  the s-numbers of the operators $X_{\varepsilon+\tau}$ and  $X_\varepsilon$   obey
\[
\sum_{j=1}^ns_j(X_{\varepsilon+\tau})\leq  \sum_{j=1}^ns_j(X_\varepsilon).
\]
\end{lemma}

{\it Proof.}
Indeed, let $P$   be the  orthogonal projection onto  the  span of  eigenvectors of $X_{\varepsilon+\tau}$ corresponding to the first $n$ eigenvalues $s_1(X_{\varepsilon+\tau}),\dots, s_n(X_{\varepsilon+\tau})$.
Then
\[
\sum_{j=1}^ns_j(X_{\varepsilon+\tau})={\rm Tr}\, \bigl(X_{\varepsilon+\tau} P\bigr)=\int_{\Bbb R}{\rm Tr}\, \bigl( PY_{\varepsilon+\tau}(\xi) P\bigr)\,d\xi=
\int_{\Bbb R}{\rm Tr}\, \bigl( PY_{\varepsilon}(\eta) P\bigr)\,d\eta.
\]
Put differently,
\[
\sum_{j=1}^ns_j(X_{\varepsilon+\tau})={\rm Tr}\, \bigl(P X_\varepsilon P\bigr)\leq \sum_{j=1}^ns_j(X_\varepsilon).
\]
$\,\Box$

Let us use induction  to prove
\begin{lemma}
Let $E_j=-\varepsilon^2_j$  be the negative eigenvalues of $\tilde H$. 
Then for each $n\in {\Bbb N}$,
\begin{equation}\label{monot}
\sum_{j=1}^n \varepsilon_j\leq   \sum_{j=1}^ns_j(X_{\varepsilon_{n}}).
\end{equation}
\end{lemma}
{\it Proof.}  Note that  \eqref{monot}  holds  for $n=1$, because $ \varepsilon_1=s_1(X_{\varepsilon_1})$.
Assume that it holds  for some $n$. Then
\[
\sum_{j=1}^n \varepsilon_j\leq  \sum_{j=1}^ns_j(X_{\varepsilon_{n}})\leq  \sum_{j=1}^ns_j(X_{\varepsilon_{n+1}}),
\]
and since $ \varepsilon_{n+1}=s_{n+1}(X_{\varepsilon_{n+1}})$, we obtain that
\[
\sum_{j=1}^{n+1} \varepsilon_j \leq  \sum_{j=1}^{n+1}s_j(X_{\varepsilon_{n+1}}),
\]
$\Box$

Theorem~\ref{T3} follows  from \eqref{tr<12V} and \eqref{monot}. $\Box$

\bigskip

Let us  now prove Theorem~\ref{T4*}.  
Traditionally,  proofs of Lieb-Thirring inequalities often start by establishing bounds for the sums of  lower powers of  eigenvalues.
 Then, these initial bounds are extended to higher powers of eigenvalues through integration techniques.
First, we need to understand the relation   between   the bottom of the spectrum of  $\tilde H$  and  the norm $\|V\|_\infty$.
The statement  below  tells us that, if $\|V\|_\infty<2$,  then  the negative eigenvalues  of $\tilde H$ are  situated to the  right of
the point
$-2\|V\|_\infty$.

\begin{proposition}\label{preced7.5} Suppose that $\|V\|_\infty<2$.
Then
\[
\tilde H+\gamma I\geq 0,
\qquad \text{for all}\qquad \gamma>2\|V\|_\infty.
\]
\end{proposition}

{\it Proof.} According to  the Birman-Schwinger principle,  it is  enough  to show that 
\begin{equation}\label{X,1as}
\|X_\varepsilon\|<1,\qquad \text{ as long as}\qquad \varepsilon^2>2\|V\|_\infty.
\end{equation}
The matrix  elements of  the operator $X_\varepsilon$ obey the  estimate
\[
\bigl|(X_\varepsilon\delta_n,\delta_m)\bigr|\leq \|V\|_\infty \frac{e^{-\varepsilon|n-m|}}{2\varepsilon}.
\]
Thus, by the Schur test,
\begin{equation}\label{normX<}
\| X_\varepsilon\|\leq \|V\|_\infty \sum_{n\in {\Bbb Z}} \frac{e^{-\varepsilon|n|}}{2\varepsilon}\leq   \|V\|_\infty \Bigl(\int_0^\infty \frac{e^{-\varepsilon x}}{\varepsilon}dx+\frac1{2\varepsilon}\Bigr)=\|V\|_\infty\bigl(\frac1{\varepsilon^2}+\frac1{2\varepsilon}\bigr).
\end{equation}
Consequently, if $\varepsilon>2$, then $\| X_\varepsilon\|<1$.  However, if $0<\varepsilon<2$, then it follows  from  \eqref{normX<}  that
\[
\|X_\varepsilon\|\leq \frac{2\|V\|_\infty} {\varepsilon^2}.
\]
This implies \eqref{X,1as}. $\Box$

\bigskip

\begin{corollary}
Let $E_j$ be the  negative  eigenvalues of $\tilde H$. Suppose that $\|V\|_\infty<2$.
Then  for  any $\gamma>0$,
\begin{equation}\label{e-g}
\sum_j (|E_j|-\gamma)^{1/2}_+\leq \sqrt2\sum_n (|V(n)|-\gamma/4)_+.
\end{equation}
\end{corollary}

{\it Proof}.  According to the preceding  proposition,   this inequality  holds  for $\gamma\geq 4$,  since in this case, the left  hand  side equals  zero. 
Assume now that $\gamma<4$. Note  that  those numbers $-(E_j+\gamma)_-$ that are  different   from zero  are the negative  eigenvalues  of  the operator
$
\tilde H +\gamma I.
$ Now
we  decompose  $V$ into the sum  $V=V_\gamma +\tilde V_\gamma$, where
\[
V_\gamma(n)=\begin{cases} 
V(n)+\gamma/4\qquad \text{if}\quad V(n)\leq- \gamma/4\\
0\qquad\quad  \text{if}\quad V(n)>-\gamma/4.
\end{cases}
\]
In this case,
$\tilde V_\gamma>-\gamma/4$ on all of ${\Bbb Z}$. Therefore, if $\gamma<4$, then
\[
-\frac12\frac{d^2}{dx^2}+\gamma I+\sum_n \tilde V_\gamma(n)\delta(x-n)\geq -\frac12\frac{d^2}{dx^2}+\gamma I-\gamma/4\sum_n \delta(x-n) \geq 0,
\]
by Proposition~\ref{preced7.5}.
Thus, we obtain the estimate
\[
\tilde H +\gamma\geq -\frac12\frac{d^2}{dx^2}+\sum_{n} V_\gamma(n)\delta(x-n).
\]
which implies that
\[
\sum_j
(|E_j|-\gamma)_+^{1/2}\leq \sqrt 2\sum_n |V_\gamma(n)|.
\]
$\Box $

\bigskip

Theorem~\ref{T4*} is a  consequence of  the stronger  result  given below:

\begin{theorem} Let $\|V\|_\infty<2$ and let $E_j$ be the negative  eigenvalues of $\tilde H$.
Then for any $p>1/2$,
\begin{equation}\label{LTp>12}
\sum_j |E_j|^p \leq C_p\sum_n|V(n)|^{p+1/2},
\end{equation}
where
\[
C_p=\frac{\sqrt 2\int_0^\infty(1-\gamma/4)_+\gamma^{p-3/2}d\gamma}{\int_0^\infty(1-\gamma)_+^{1/2}\gamma^{p-3/2}d\gamma}.
\]
\end{theorem}

{\it Proof}. It is  enough to multiply both  sides  of \eqref{e-g}   by $\gamma^{p-3/2}$  and integrate  the resulting  functions   with respect to $\gamma$ from $0$ to $\infty$. 
Making the  substitution   $\gamma=|E_j|\tilde \gamma$  in each integral  term of the  sum,  we obtain
\begin{equation}\label{g=eg}
\sum_j  \int_0^\infty (|E_j|-\gamma)^{1/2}_+\gamma^{p-3/2}d\gamma=\sum_j |E_j|^p \int_0^\infty (1-\gamma)^{1/2}_+\gamma^{p-3/2}d\gamma
\end{equation}
Similarly,
\begin{equation}\label{g=Vg}
 \sum_n \int_0^\infty(|V(n)|-\gamma/4)_+\gamma^{p-3/2}d\gamma=\sum_n |V(n)|^{p+1/2}\int_0^\infty(1-\gamma/4)_+\gamma^{p-3/2}d\gamma.\end{equation}
The estimate \eqref{LTp>12}  follows  now  from \eqref{e-g},  \eqref{g=eg}, and \eqref{g=Vg}.   $\Box$

\section{Proofs  of Theorems~\ref{alpha<1/2},  ~\ref{alpha=1/2} and ~\ref{randomac}}

Some of our arguments rely on  the subordination (or subordinacy) theory.   
That  is a technique developed by Gilbert and Pearson (see \cite{G} and \cite{GP}) to analyze the spectrum of Schr\"odinger operators,   It establishes a correspondence between the spectral properties of an operator and the behavior of solutions to an associated eigenvalue equation,  identifying the singular  and absolutely continuous parts of the spectrum.
First of all,  we  note
that a formal solution to
\[
H_\omega u= k^2 u
\]  is a function whose  values at integer points $n\in {\Bbb Z}$
obey the condition
\begin{equation}\label{snova}
-u(n+1)-u(n-1)+2\cos(k) u(n)+\frac{\sin(k)}{k}\bigl(a+V_\omega(n)\bigr)u(n).
\end{equation}
Therefore,  the study of  the operator $H_\omega$ is reduced  to the  study
of the  equation of the   form
\begin{equation}\label{snova1}
-u(n+1)-u(n-1)+W(n)\, u(n)=E u(n)
\end{equation} with an appropriate real-valued potential $W$ the  choice  of which  depends on $k$.

Let  $u$  and $v$  be two non-zero solutions  to \eqref{snova1}.
Define the family of thr norms $\|\cdot\|_L$  by
\[
\|u\|^2_L=\sum_{n=1}^{[L]} |u(n)|^2+(L-[L])|u([L]+1)|^2,
\]
where $[L]$  denotes the integer part of $L$.
We will say that $v$ is a   subordinate  solution  to \eqref{snova1} provided
\begin{equation}\label{u,v}
\lim_{L\to\infty}\frac{\|v\|_L}{\|u\|_L}=0,
\end{equation}
for any other linearly independent solution $u$.

The collection of points  $E$   for which \eqref{snova1} does not have a  subordinate solution  is an essential support of
the absolutely  continuous spectrum of  the operator $\tilde H$
defined  on $\ell^2({\Bbb N})$ by
\[
[\tilde H u](n)=-u(n+1)-u(n-1)+W(n)\, u(n),\qquad \text{with}\quad u(0)=0.
\]
A  similar statement  holds  for  continuous Schr\"odinger operators  on ${\Bbb R}_+$ with $\|u\|_L$
defined by
\[
\|u\|_L^2=\int_0^L|u(x)|^2dx.
\]

\bigskip

{\it Proof  of Theorem~\ref{alpha=1/2}}.
We  apply Lemma 8.8  from  the paper \cite{KLS} by Kiselev,  Last and Simon.   
Define 
\begin{equation}\label{betagamma}
\beta=\frac{\varkappa \sin(k)}{k} \qquad \text{and}\quad \gamma=\frac{a \sin(k)}{k}  +2\cos(k).
\end{equation}
Observe that if $k>0$, then $k^2\in \sigma_{\rm ess}(H_\omega)$  if and only if $|\gamma|\leq 2$. Therefore,  it follows from
Lemma 8.8  of   the paper \cite{KLS} that,    if $\lambda=k^2>0$ belongs to $\sigma_{\rm ess}(H_\omega)$ and $\gamma\neq \pm 2,\pm \sqrt 2, 0$, then there is
a  solution  to  \eqref{snova}  that decays  at positive infinity as $O(n^{-p})$ with
\begin{equation}\label{pdefinition}
p=\frac{\beta^2}{8-2\gamma^2}.
\end{equation}
More precisely,  this  solution $v$ behaves asymptotically as 
\[
v(n)\sim n^{-p},\qquad \text{for}\quad n\to \infty.
\]
Clearly,  $v\in \ell^2({\Bbb N})$     if  and only if  $p>1/2$.
Thus,  by the general argument of the rank one preturbation theory,  the region where  $p>1/2$  contains  only pure point spectrum.

It is easy to  see that   for  any $|\gamma|<2$,  the   function $v$  is  a subordinate  solution  of \eqref{snova}.
Indeed, for any other solution $u$ of \eqref{snova}, the Wronskian
\[
{\mathcal W}[u,v]=u(n+1)v(n)-u(n)v(n+1)
\]
does not depend on $n\in {\Bbb N}$.  Consequently,  there is a  constant $C>0$  independent of $n$
for which
\[
|u(n)|+|u(n+1)|\geq C n^{p},\qquad \forall n\in {\Bbb N}.
\]
Thus \eqref{u,v}  holds, which proves that $v$  is  subordinate.

This implies that the region where $|\gamma|<2$ is free of the absolutely continuous  spectrum.
Consequently, if $|\gamma|<2$  and $p\leq 1/2$, then $k^2$  belongs to the  singular continuous spectrum.

The negative points of the  essential spectrum could  be  analyzed in a similar way.  One only needs to replace the functions $\sin$ and $\cos$ by $\sinh$ and $\cosh$.
$\Box$

\bigskip

{\it Proof of Theorem~\ref{randomac}}.  Consider the case $a>0$. The  proof in the  case $a<0$ is   similar.  Let $\gamma$ be defined by \eqref{betagamma}.
Let $u$ be the solution of  the equation \eqref{snova}.    For each  $k^2\in \sigma_{\rm ess}(H_\omega)$, let  $\tilde k\in [0,\pi]$  be the unique solution of   the equation 
\[ 
2\cos k+\frac{a \sin k}{k}=2\cos \tilde k.
\]
Define the  functions $R(n)$  and $\theta(n)$  by
\[
\begin{split}
R(n)\cos(\theta(n))=u(n)-\cos(\tilde k) u(n-1),\\
R(n) \sin(\theta(n))=\sin( \tilde k) u(n-1).
\end{split}
\]
The ambiguity in the definition of $\theta$  is  resolved by the condition  $\theta(n+1)-\theta(n)\in [-\pi,\pi)$.
Then according to the  formulas (2.12a)-(2.12c) of the paper \cite{KLS},
\[
\frac{R(n+1)^2}{R(n)^2}=1+\frac{\sin k}{k\sin \tilde k}V_\omega(n)\sin(2(\theta(n)+\tilde k))+\frac{\sin^2 k}{k^2\sin^2 \tilde k}V^2_\omega(n)\sin^2(\theta(n)+\tilde k)
\]
while
\[
\cot(\theta(n+1))=\cot(\theta(n)+\tilde k)+\frac{\sin k}{k\sin \tilde k}V_\omega(n).
\]
In particular, we see that $R(n)$ and $\theta(n)$   depend only on  the  random variables $\omega_j$ with $j\leq n-1$.
Therefore,   since
\[
R(n+1)^4=\Bigl(1+\frac{\sin k}{k\sin \tilde k}V_\omega(n)\sin(2(\theta(n)+\tilde k))+\frac{\sin^2 k}{k^2\sin^2 \tilde k}V^2_\omega(n)\sin^2(\theta(n)+\tilde k)\Bigr)^2R(n)^4,
\]
we  conclude that
\[
{\Bbb E}(R(n+1)^4)\leq  \Bigl(1+\frac{3\varkappa^2\sin^2 k}{k^2\sin^2 \tilde k}n^{-2\alpha}+\frac{\varkappa^4\sin^4 k}{k^4\sin^4 \tilde k}n^{-4\alpha}\Bigr){\Bbb E}(R(n)^4).
\]
Here ${\Bbb E}(\cdot)$  denotes  the expectation.  The  last inequality implies
that
\[
\log\bigl({\Bbb E}(R(n)^4)\bigr)\leq \Bigl(C_1 \frac{\varkappa^2\sin^2 k}{k^2\sin^2 \tilde k}+C_2\frac{\varkappa^4\sin^4 k}{k^4\sin^4 \tilde k}\Bigr)\log\bigl({\Bbb E} (R(1)^4)\bigr)
\]
with positive  constants $C_1$ and $C_2$ depending only on $\alpha$.  Since $R(1)$  does not depend on $\omega$, it could be interpreted as  a constant  that we  choose.
Let  now $I$  be a   closed bounded interval  that is contained  in the interior of  one band of  the  essential spectrum of $H_\omega$.
Then $\min_{k^2\in I} |\sin  \tilde k|>0$. Therefore,
\[
{\Bbb E} \bigl(\int_I  R(n)^4  \, dE \bigr)=
\int_I {\Bbb E} (R(n)^4) \, dE<C<\infty
\]
where the constant $C$ depends only on  the interval $I$, the choice of $R(1)$, and the values of $\alpha$  and $\varkappa$.
Thus,  by Fatou's lemma,  we infer   from this inequality   that 
\begin{equation}\label{Fatoulemma}
 \liminf_{n\to \infty}\int_I  R(n)^4  \, dE<\infty
\end{equation}
for almost every $\omega$.
Now we apply Theorem 1.3  from \cite{KLS} according to which condition \eqref{Fatoulemma} implies 
that  the  spectral measure of $H_\omega$  is absolutely continous on the interior  of  the interval $I$.
$\,\,\Box$

\bigskip

{\it  Proof  of Theorem~\ref{alpha<1/2}}  Again, we consider only the case $a>0$.
Let $p$  be defined  by \eqref{pdefinition}  with $\beta$ and $\gamma$  defined by \eqref{betagamma}.
The second line of the short  proof  of Theorem 8.6  from \cite{KLS} tells us  that, if $|\gamma|<2$ there is a  solution to \eqref{snova}
decaying at infinity as $\exp(-\tau |n|^{1-2\alpha})$  with
\[
\tau=(1-2\alpha)p.
\]
This solution is $\ell^2$. By the  general principles of the rank one perturbation theory,    the region  where $|\gamma|<2$  contains only pure point spectrum.
$\,\,\Box$

\end{document}